\documentclass[pra,aps,showpacs,amssymb,amsmath,amstext,amsfont,noshowkeys,groupedaddress,12pt,nofootinbib]{revtex4}
\usepackage{graphicx}
\usepackage{accents}
\usepackage{enumerate}

\newtheorem{theorem}{Theorem}

 \numberwithin{equation}{section}

\begin{document}
\title{Symmetries of Cosmological Cauchy Horizons\\ with Non-Closed Orbits}
\author{Vincent Moncrief}
\affiliation{Department of Physics and Department of Mathematics, \\ Yale University, P.O. Box 208120, New Haven, CT 06520, USA. \\ E-mail address: vincent.moncrief@yale.edu}
\author{James Isenberg}
\affiliation{Department of Mathematics,\\ University of Oregon, Eugene, OR 97403, USA. \\ E-mail address: isenberg@uoregon.edu}

\date{\today}
\begin{abstract}
We consider analytic, vacuum spacetimes that admit compact, non-degenerate Cauchy horizons. Many years ago we proved that, if the null geodesic generators of such a horizon were all \textit{closed} curves, then the enveloping spacetime would necessarily admit a non-trivial, horizon-generating Killing vector field. Using a slightly extended version of the Cauchy-Kowaleski theorem one could establish the existence of infinite dimensional, analytic families of such `generalized Taub-NUT' spacetimes and show that, generically, they admitted \textit{only} the single (horizon-generating) Killing field alluded to above. In this article we relax the closure assumption and analyze vacuum spacetimes in which the generic horizon generating null geodesic densely fills a 2-torus lying in the horizon. In particular we show that, aside from some highly exceptional cases that we refer to as `ergodic', the non-closed generators always have this (densely 2-torus-filling) geometrical property in the analytic setting.

By extending arguments we gave previously for the characterization of the Killing symmetries of higher dimensional, stationary black holes we prove that analytic, 4-dimensional, vacuum spacetimes with such (non-ergodic) compact Cauchy horizons always admit (at least) two independent, commuting Killing vector fields of which a special linear combination is horizon generating. We also discuss the \textit{conjectures} that every such spacetime with an \textit{ergodic} horizon is trivially constructable from the flat Kasner solution by making certain `irrational' toroidal compactifications and that degenerate compact Cauchy horizons do not exist in the analytic case.
\end{abstract}
\maketitle

\section{Introduction}
\label{sec:introduction}
To disprove the \textit{cosmic censorship conjecture} it would suffice to establish the existence (in a suitable function space topology) of an open set of globally hyperbolic solutions to the vacuum Einstein equations which are each extendible, through Cauchy horizons, beyond their maximal Cauchy developments. Analytic examples of such extendible spacetimes include the Taub metric on \(\mathbf{S}^3 \times \mathbb{R}\) and the flat Kasner metric on \(T^3 \times \mathbb{R}\). Each of these solutions can be (analytically) extended through a compact Cauchy horizon to include an acausal region containing closed timelike curves. If this feature were actually stable against sufficiently small perturbations then cosmic censorship would be false.

To study this stability question, within the convenient framework of (real-)analytic metrics, one can employ a straightforward generalization of the Cauchy-Kowalewski theorem to prove the existence of infinite dimensional families of `generalized Taub-NUT' vacuum spacetimes, with a variety of spatial topologies, which each, as in the examples mentioned above, contain a compact Cauchy horizon separating globally hyperbolic and acausal regions \cite{Moncrief:1984,Moncrief:1982}. These families, large though they are, fail to disprove cosmic censorship for several reasons.

First of all every such generalized Taub-NUT solution admits at least one Killing vector field---a vector which is spacelike in the globally hyperbolic region, null on the Cauchy horizon (and hence tangent to the horizon's null geodesic generators) and timelike in the acausal extension. Thus these particular families could not possibly fill (even  densely) an open subset of generically non-symmetric solutions in any reasonable function space topology. Secondly, even within the circumscribed context of analytic metrics admitting at least one Killing field they require a further special restriction upon their `initial data' (which, by exploiting analyticity and the extended Cauchy-Kowalewski theorem can be specified on the horizon itself) which, roughly speaking, corresponds to a Lagrangian submanifold of the full set of solutions of the chosen (one-Killing-field) symmetry type. To rigorously treat the complementary family of one-Killing field metrics (i.e., to relax the Lagrangian submanifold restriction) has necessitated a still further generalization of the Cauchy-Kowalewski theorem through the development of so-called Fuchsian methods \cite{Isenberg:2002,Choquet-Bruhat:2004,Choquet-Bruhat:2006} but the spacetimes obtained by these techniques typically exhibit strong curvature singularities instead of Cauchy horizons and so are inextendable beyond their maximal Cauchy developments. Finally the generalized-Taub-NUT solutions are all (real) analytic which many might regard as an artificial restriction to place on any supposedly physically relevant family of vacuum spacetimes.

Since the presence of a Killing field seemed to play a crucial role in the construction of these generalized Taub-NUT spacetimes it is of interest to ask whether perhaps the occurrence of such a field was in fact \textit{necessary} for the existence of a compact Cauchy horizon, at least in the (vacuum) analytic case. In earlier articles \cite{Moncrief:1983,Isenberg:1985,Isenberg:1992} we showed that this was indeed the case provided that the null-generating geodesic curves which foliate the horizon are all closed. While this might at first seem to be an unduly artificial restriction upon the geometry of the horizon we now believe that it represents the least constraining assumption and that the failure of all the null generators to be closed implies the existence of at least a second Killing field. By contrast the known (analytic) solutions with all closed generators need only have the single Killing field which is tangent to the horizon's generators.

In this paper we prove, under certain assumptions, that the occurrence of an (analytic) compact Cauchy horizon with non-closed generators implies the existence of at least one Killing field---always tangent to the horizon's generators---and we have already shown elsewhere that the presence of such a Killing field with non-closed integral curves implies the existence of a second Killing field \cite{Isenberg:1992}. We know of examples (see below) in which even a third Killing field is required by the special nature of the geometry but we do not have a systematic treatment of this case which are refer to as `ergodic'.

The main assumption we need, in addition to analyticity and the imposition of the vacuum field equations is that the compact Cauchy horizon be non-degenerate in the sense that at least one (and hence, as we prove, every in the case of a connected horizon) of its null geodesic generators be incomplete in one direction. In fact we do not know of an example of a degenerate Cauchy horizon (though compact, degenerate null hypersurfaces which are not Cauchy horizons can certainly exist for (electro-)vacuum spacetimes) and, in the case of closed generators we could even prove their non-existence on certain topologies. We suspect that degenerate compact Cauchy horizons may not exist in general for analytic (electro-)vacuum spacetimes  but do not have a proof of this surmise. The second assumption we require is that the horizon be non-ergodic in the sense that it not be densely filled by the orbit of any single geodesic generator. Examples of vacuum spacetimes with ergodic Cauchy horizons do exist and can be created from the flat Kasner metric through spatial compactification with an `irrational' shift in the  obvious identifications to produce a toroidal horizon which each null generator densely fills. We suspect that, up to finite covers, these solutions (which have the extra, third Killing field alluded to above) may exhaust the vacuum ergodic horizon cases but also have no proof of this conjecture. On the other hand the ergodic case could, to some extent, be treated by a straightforward generalization of the techniques developed here provided that the assumed compact Cauchy horizon admits an analytic foliation with compact (2-dimensional) leaves, transversal to the given the null geodesic `flow'. While we also impose the vacuum field equations it seems quite likely that our results could be readily generalized to allow for certain types of matter sources. Indeed the original results for closed generators were derived for the electro-vacuum field equations.

Analyticity is the final restrictive assumption that we make but this hypothesis has a certain double-edged quality that makes it seem less objectionable than it appears at first sight. First of all, if a genuine open set (in some suitable function space topology) of vacuum spacetimes admitting compact Cauchy horizons did exist it would presumably contain a large (perhaps densely filling) subset of analytic solutions. Thus one could expect to probe such a set by focusing on its analytic elements. Secondly, analyticity serves, by its very rigidity, to exclude the occurrence of many exotic types of cosmological boundaries which could otherwise occur through suitable (non-analytic) `fine-tuning' of the `initial data'. For example in the special case of polarized Gowdy metrics on \(T^3 \times \mathbb{R}\) one can exploit non-analyticity to produce a large variety of, highly non-generic, cosmological boundaries involving such exotica as Kantor sets of curvature singular regions interspersed with complementary sets of non-singular Cauchy horizon \cite{Chrusciel:1990}. The fine-tuning of the data needed to produce these exotica is incompatible with analyticity so that, in concentrating on analytic solutions, one avoids being distracted by such mathematically allowed but non-generic features. Any truly generic feature should survive analytic approximations. Thus analyticity is actually an advantage rather than a liability if only stable properties are of interest.

The main difficulty in treating the problem of non-closed generators considered here, over and above those already handled for the closed generator case, is a proof that the candidate vector field for the horizon generating Killing field is in fact analytic. Otherwise much of the argument goes through in essentially the same way as for the closed generator case. The hypothetical Killing field, restricted to the horizon, is everywhere parallel to the generators and so already determined up to a multiplicative factor. We define this factor (in the non-degenerate case wherein every generator is incomplete to the future) by the requirement that the future affine length of every null geodesic generator be a fixed positive number \(2/k\) provided one takes the initial condition for the generator starting at an arbitrary point \textit{p} of the horizon to have its tangent vector given by the hypothetical Killing field \(X(p)\) at that point. In other words one adjusts the multiplicative factor until each generator (taken with these rescaled initial conditions) has future length \(2/k\). The technical problem is then to prove that the needed rescaling factor is in fact analytic. In the closed generator case we found an explicit formula for this factor from which its analyticity was apparent but here we seem to need a more subtle argument involving the convergence of a sequence of analytic approximations to the needed rescaling factor. Unfortunately though since real analytic functions do not form a nice Banach space (with the norm of uniform convergence) we have had to `artificially' complexify the analytic structure of the horizon and carry out the convergence argument in the complexified context, extracting the desired analyticity of the real section at the end of this analysis. While workable this complicating feature is rather disappointing in comparison with the simplicity of the corresponding closed generator argument and so one wonders whether perhaps a further simplification could be found for the present problem.

Our results have some natural correspondences with those for the (Killing) event horizons of stationary black holes and one can compactify these latter horizons to obtain examples (in certain cases) of `cosmological' compact Cauchy horizons of the sort we are interested in. In the black hole case, for which there is a natural normalization of the Killing horizon generator, the constant \textit{k} is essentially the so-called surface gravity of the horizon \cite{Moncrief:2008}. It might seem that one could produce examples of degenerate Cauchy horizons (having, by definition, \(k = 0\)) by compactifying the event horizons of extreme black holes. The simplest (electro-vacuum) example however is provided by the extreme Reissner-Nordstrom metric with horizon generating Killing field given, in standard coordinates, by \(\frac{\partial}{\partial t}\). We can compactify the horizon at \(r = r_+ = M\) to \(\mathbf{S}^2 \times \mathbf{S}^1\) by identifying the points labeled \(\lbrace t, \theta, \varphi\rbrace\) with those labeled \(\lbrace t + \ell, \theta, \varphi\rbrace\) for a fixed constant \(\ell \neq 0\). However the extreme black hole metric has \(\frac{\partial}{\partial t} \cdot \frac{\partial}{\partial t} = -\left( 1 - \frac{M}{r}\right)^2\) so that the generating vector field \(\frac{\partial}{\partial t}\), tangent to the \(\mathbf{S}^1\) fibers, has closed timelike orbits on both sides of the compact null surface at \(r = M\) which can therefore not be a Cauchy horizon. A similar phenomenon occurs for the more general extreme Kerr-Newman solution.

Though the Killing field or fields we produce via the extended Cauchy-Kowalewski theorem are possibly only determined by convergent expansions in some neighborhood of the assumed Cauchy horizon it is straightforward to show that these automatically propagate (as solutions of Killing's equation) to the full maximal Cauchy development on the globally hyperbolic side of the horizon. This follows from the well-known fact that in (for simplicity) a vacuum spacetime any Killing field satisfies a linear hyperbolic equation which in fact preserves the vanishing of the Killing form for the propagated vector field \cite{Moncrief:1983,Coll:1977}.  
\section{Construction of the Candidate Vector Field}
\label{sec:construction}

\subsection{Geometrical Assumptions and Basic Constructions}
\label{subsec:geometrical}
We shall be considering real analytic, time orientable, vacuum spacetimes \(({}^{(4)}\!V,g)\) which contain compact Cauchy horizons. More precisely, we assume that \({}^{(4)}\!V = M \times \mathbb{R}\), where \textit{M} is a compact, connected, analytic and orientable three-manifold without boundary, and that \textit{g} is an analytic, Lorentzian, Ricci-flat metric on \({}^{(4)}\!V\). We also assume that \(({}^{(4)}\!V,g)\) admits a compact, embedded null hypersurface \textit{N}, which can be realized as a level surface of some real analytic function \(\tau\) with no critical points on a neighborhood of \textit{N}, and that \textit{N} is a Cauchy horizon for one of the two open submanifolds of \({}^{(4)}\!V\) which \textit{N} separates. Thus we regard \({}^{(4)}\!V\) as a disjoint union \({}^{(4)}\!V_+ \cup N \cup {}^{(4)}\!V_-\) where \({}^{(4)}\!V_\pm = M \times \mathbb{R}_\pm\) (with \(\mathbb{R}_\pm = \lbrace r \gtrless 0\rbrace\) and assume that at least one of the two spacetimes \(({}^{(4)}\!V_+,g_+), ({}^{(4)}\!V_-,g_-)\) (where \(g_+\) and \(g_-\) represent the restriction of \textit{g} to \({}^{(4)}\!V_+\) and \({}^{(4)}\!V_-\) respectively) is globally hyperbolic. For convenience, we may assume that the function \(\tau\) has been chosen so that \textit{N} coincides with the level surface of \(\tau\) having the level value \(\tau = 0\).

Since \textit{N} is null and since by assumption \(\tau\) has no critical points on a neighborhood of \textit{N}, the vector field \({}^{(4)}\!X\) determined by \(d\tau\) (i.e. given in local charts by \({}^{(4)}\!X^\alpha = g^{\alpha\beta} \tau_{,\beta}\)) is non-vanishing on a neighborhood of \textit{N}, null on the surface \textit{N} and thus tangent to the null geodesic generators of that surface. Let \textit{X} designate the restriction of \({}^{(4)}\!X\) to the null surface \textit{N} so that \textit{X} may be viewed as a vector field defined on \textit{N} itself.

Since \textit{X} is non-vanishing and tangent to the null geodesic generators of \textit{N}, one can always choose local coordinates \(\lbrace x^a, x^3\rbrace\) on suitable open subsets of \textit{N} such that the \(\lbrace x^a\; | a = 1,2\rbrace\) are constant along the null generators and such that \(X = \frac{\partial}{\partial x^3}\) within each such local chart. One can construct such charts in the following way. Choose a two-disk \textit{D} which is (analytically) embedded in \textit{N} and transversal to the flow of \textit{X} and let \(\lbrace x^a\rbrace\) be coordinates on \textit{D}. Define coordinates \(\lbrace x^a, x^3\rbrace\) on a tubular neighborhood \(\approx D \times I\) of \textit{D} in \textit{N} by requiring that the \(x^a\) remain constant along the integral curves of \textit{X} and that \(x^3\) coincide with the natural integral curve parameter determined by \textit{X} (after fixing, say, \(x^3\; |_D = k(x^a)\) for some real analytic function \textit{k} defined on \textit{D}). The range of \(x^3\) may, for convenience, be allowed to vary from generator to generator with, for example, \(k(x^a) - \delta_-(x^a) < x^3 < k(x^a) + \delta_+(x^a)\) where \(\delta_\pm\) are two strictly positive real analytic functions. On the connected components of the domains of intersection of any two such local charts, the two sets of coordinate functions \(\lbrace x^a, x^3\rbrace, \lbrace x^{a'}, x^{3'}\rbrace\) are clearly related by a transformation of the form
\begin{equation}\label{eq:201}
\begin{split}
x^{3'} &= x^3 + h(x^a)\\
x^{a'} &= x^{a'} (x^b)
\end{split}
\end{equation}
where \textit{h} is an analytic function and \(x^{a'}(x^b)\) a local (analytic) diffeomorphism defined on some transversal two manifold which lies in the domains of both charts.

We shall often consider local charts of the type described above not only for the fixed vector field \textit{X} but also for other analytic vector fields defined on \textit{N} which are tangent to its null generators. If \textit{K} is some non-vanishing vector field on \textit{N} tangent to the generators of \textit{N} and we set up local charts of the type described above based on \textit{K} (rather than on \textit{X}), then the connected components of the domains of intersection of the new charts (say, \(\lbrace x^{3'}, x^{a'}\rbrace\) with \(K = \frac{\partial}{\partial x^{3'}}\)) with the old ones \(\lbrace x^3, x^a\rbrace\) for which \(X = \frac{\partial}{\partial x^3}\) necessarily admit coordinate transformations of the form
\begin{equation}\label{eq:202}
\begin{split}
x^{3'} &= h(x^3, x^a)\\
x^{a'} &= x^{a'}(x^b)
\end{split}
\end{equation}
where, as before, \(x^{a'}(x^b)\) is a local diffeomorphism and where \(\frac{\partial h}{\partial x^3} \neq 0\).

Let \(\lbrace x^a, x^3\rbrace\) be local coordinates of the type described above defined on some domain \(U \approx D \times I\) lying in \textit{N} and adapted to some fixed non-vanishing vector field \textit{K} (i.e., chosen so that \(K = \frac{\partial}{\partial x^3}\) within the chart) which is tangent to the null generators of \textit{N}. Then one can always construct a local chart \(\lbrace t, x^a, x^3\rbrace\) on some domain \({}^{(4)}\!U\) of \({}^{(4)}\!V\) which intersects \textit{N} in \textit{U}, for which the hypersurface \(U = N \cap {}^{(4)}\!U\) corresponds to the level value \(t = 0\) and in terms of which the Lorentzian metric \textit{g} takes the convenient form
\begin{equation}\label{eq:203}
\begin{split}
g &= dt \otimes dx^3 + dx^3 \otimes dt\\
 &{} + \varphi\; dx^3 \otimes dx^3 + \beta_a (dx^a \otimes dx^3 + dx^3 \otimes dx^a)\\
 &{} + \mu_{ab}\; dx^a \otimes dx^b.
\end{split}
\end{equation}

By construction, the coordinates, restricted to the null surface \(t = 0\), coincide with those of the original chart defined on \textit{U} and because \textit{N} is null and \textit{g} is Lorentzian, the metric functions obey
\begin{equation}\label{eq:204}
\varphi|_{t=0} = \beta_a|_{t=0} = 0,
\end{equation}
with \(\mu_{ab}\) pointwise positive definite (as a \(2 \times 2\) symmetric matrix). The construction of such local charts on suitable domains in (\({}^{(4)}\!V,g\)) was discussed in detail in section II B of Ref.~\cite{Moncrief:1983} and need not be repeated here. The local (analytic) coordinate functions \(\lbrace t, x^a, x^3\rbrace\) are uniquely determined by the local chart \(\lbrace x^a, x^3\rbrace\) defined on \(U \subset N\) and by the coordinate conditions implicit in the desired metric form (\ref{eq:203}).

Because of their resemblance to gaussian normal coordinates (but with \(\frac{\partial}{\partial t}\) tangent to \textit{null} geodesics transversal to \textit{N} instead of timelike ones), we called the coordinate systems for which \textit{g} takes the form (\ref{eq:203}) and satisfies (\ref{eq:204}) \textit{gaussian null} coordinates. In the present context, when we wish to emphasize that the coordinates have, in addition, been \textit{adapted} to some particular vector field tangent to the generators of \textit{N} (i.e., chosen so that \(\left.\frac{\partial}{\partial x^3}\right|_{t=0}\) coincides with the given vector field) we shall refer to them as \textit{adapted gaussian null} coordinates or \textit{agn} coordinates for brevity.

The Einstein equations are written out in detail in an arbitrary gaussian null coordinate chart in Section II C of Ref.~\cite{Moncrief:1983}. As in that reference, we shall often use the notation of an overhead nought to signify restriction to the null surface \textit{N} (labeled in gaussian null coordinates by \(t = 0\)). Thus, for example, we shall often write \(\mathring{\mu}_{ab}\) for \(\mu_{ab}|_{t=0}\), etc., and can therefore reexpress Eqs.~(\ref{eq:204}) as \(\mathring{\varphi} = 0, \mathring{\beta}_a = 0\).

\subsection{Invariance of the Transversal Metric}
\label{subsec:invariance}
Consider an arbitrary two-disk \textit{D} which is analytically embedded in \textit{N} and which is everywhere transversal to the null generators of that hypersurface. In a gaussian null coordinate chart which covers \textit{D}, it is clear that \textit{D} has a coordinate characterization of the form,
\begin{equation}\label{eq:205}
t = 0,\; x^3 = f(x^a)
\end{equation}
for some real analytic function \textit{f}. (Here the \(\lbrace x^a\rbrace\) range over those values corresponding to the generators which intercept \textit{D}.) From Eqs.~(\ref{eq:203}), (\ref{eq:204}) and (\ref{eq:205}) one sees that \textit{g} induces a Riemannian metric \(\mu_D\), given by
\begin{equation}\label{eq:206}
\mu_D = \mu_{ab}|_{x^3=f(x^c)} dx^a \otimes dx^b
\end{equation}
on \textit{D}. If we let \textit{D} flow along the integral curves of the vector field \(K = \frac{\partial}{\partial x^3}\) associated (at least locally) to the chosen chart, then we get a one-parameter family \(D_\lambda\) of embeddings of \textit{D} in \textit{N} characterized by
\begin{equation}\label{eq:207}
t = 0,\; x^3 = f(x^a) + \lambda
\end{equation}
and a corresponding family of metrics \(\mu_{D_\lambda}\) given by
\begin{equation}\label{eq:208}
\mu_{D_\lambda} = \mu_{ab}|_{x^3=f(x^c)+\lambda} dx^a \otimes dx^b.
\end{equation}
Here \(\lambda\) ranges over some open interval containing \(\lambda = 0\).

Locally one can always choose a particular vector field \textit{K} tangent to the null generators of \textit{N} such that the integral curves of \textit{K} coincide with the \textit{affinely parametrized} null geodesics generating \textit{N} (i.e., such that the curves \(\lbrace x^a(\lambda)\rbrace\) defined by \(t(\lambda) = 0, x^a(\lambda) = \hbox{constant}, x^3(\lambda) = \mathring{x}^3 + \lambda\) are affinely parametrized null geodesics generating (a portion of) \textit{N}, with \(\lambda\) an affine parameter). \textit{K} is of course not unique (since there is no canonical normalization for \(\lambda\) along each generator) but can be fixed by prescribing it at each point of some transversal two-manifold. In general, \textit{K} may also not be extendable to a globally defined vector field on \textit{N} (since the affinely parametrized generators of \textit{N} may be incomplete wheras the flow of a globally defined vector field on the \textit{compact} manifold \textit{N} must be complete) but this is of no consequence in the following construction. For any point \(p \in N\) choose a disk \textit{D} which contains \textit{p} and is everywhere transversal to the null generators of \textit{N}. Construct, on a neighborhood of \textit{D} in \textit{N}, a vector field \textit{K} of the type described above and let \(\lbrace x^a\rbrace = \lbrace t, x^3, x^a\rbrace\) be an agn coordinate chart adapted to \textit{K} (i.e., so that \(\frac{\partial}{\partial x^3} = K\) is thus tangent to the affinely parametrized generators of \textit{N}). Now let \textit{D} flow along the integral curves of \textit{K} to get a one-parameter family of embedded disks \(D_\lambda\) and a corresponding family of induced Riemannian metrics \(\mu_{D_\lambda}\) as described above.

In terms of this construction, one can compute the \textit{expansion} \(\hat{\theta}\) of the null generators at \textit{p} by evaluating
\begin{equation}\label{eq:209}
\hat{\theta}(p) = \left.\left(\frac{\partial}{\partial\lambda} \ell n{(\det{\mu_{D_\lambda}})}\right)\right|_{\substack{\lambda=\lambda(p)\\
x^a=z^a(p)}}.
\end{equation}
It is not difficult to verify that this definition is independent of the particular choice of transversal manifold \textit{D} chosen through \textit{p} and of the particular coordinates \(\lbrace x^a\rbrace\) used to label the generators near \textit{p}. In fact, this definition of \(\hat{\theta}\) is equivalent to the usual definition of the \textit{expansion} of the null generators of a null hypersurface.

In our case, however, \textit{N} is not an arbitrary null surface. It is, by assumption a compact Cauchy horizon in a vacuum spacetime. For such a hypersurface Hawking and Ellis have proven the important result that \(\hat{\theta}\) vanishes at every point \(p \in N\) \cite{Hawking:1973,Hollands:2007}. Thus in an agn coordinate chart adapted to \textit{K} one has
\begin{equation}\label{eq:210}
\left.(\det{\mu_{ab}})_{,3}\right|_{t=0} = 0
\end{equation}
at every point of \textit{N} covered by the chart. However, the Einstein equation \(R_{33} = 0\), restricted to \textit{N}, yields
\begin{equation}\label{eq:211}
\begin{split}
\mathring{R}_{33} &= 0 = \left\lbrack\vphantom{\frac{1}{4}} \left(\ell n \sqrt{\det{\mu}}\right)_{,33}\right.\\
&{} + \frac{1}{2} \varphi_{,t} \left(\ell n \sqrt{\det{\mu}}\right)_{,3}\\
&{} \left.\left. + \frac{1}{4} \mu^{ac} \mu^{bd} \mu_{ab,3} \mu_{cd,3}\right\rbrack\right|_{t=0}
\end{split}
\end{equation}
in an arbitrary gaussian null coordinate chart (where \((\det{\mu}) \equiv \det{(\mu_{ab})}\)). Combining Eqs.~(\ref{eq:210}) and (\ref{eq:211}), we see that \(\mu_{ab,3}|_{t=0} = 0\) throughout the local chart adapted to \textit{K}.

From this result, it follows easily that the metric \(\mu_D\) induced upon an arbitrary disk transversal to a given bundle of null generators of \textit{N} is, in fact, independent of the disk chosen. To see this one computes, recalling Eqs.~(\ref{eq:203}) and (\ref{eq:204}), the metric induced upon an arbitrary such disk \textit{D} (satisfying \(t = 0, x^3 = f(x^a)\)). From the result that \(\mu_{ab,3}|_{t=0} = 0\) it follows that this induced metric is independent of the function \textit{f} (which embeds \textit{D} in the given bundle) and hence of the particular transversal disk chosen. Though this calculation was carried out using a special family of charts, the definition of the induced metric is a geometrical one and thus the invariance of this metric (relative to an arbitrary displacement along the null generators of \textit{N}) is independent of any choice of charts.

The invariance of this transversal metric will play an important role in the sections to follow. Notice that if one starts at a transversal disk \textit{D} and flows along the generators of \textit{N}, then one may eventually reach another disk \(D'\) transversal to the same bundle of generators which partially or completely coincides with \textit{D}. Indeed, upon application of the {Poincar\'{e}} recurrence theorem in the next subsection, we shall see that this always happens and that every null generator of \textit{N} is either closed or comes arbitrarily close to closing. By the result of the preceding paragraph, the metric \(\mu_{D'}\) induced on \(D'\) is isometric to the metric \(\mu_D\) induced upon \textit{D} (since the transversal metric is invariant under the flow which carries \textit{D} to \(D'\)). If the null generators intersecting \textit{D} were all closed curves this would hardly be surprising since \textit{D} would eventually coincide with \(D'\) and the isometry would simply be the identity map. In the non-closed case of primary interest here, however it leads to non-trivial restrictions upon the transversal metric \(\mu_D\). For example, suppose \(U \subset D\) is an open subset of \textit{D} which, upon translation along the generators of \textit{N}, reintersects \textit{D} in another open set \(U'\). There is a natural diffeomorphism \(\varphi_U\) of \textit{U} and \(U'\) defined by this translation mapping and, from the invariance of the transversal metric, it follows that
\begin{equation}\label{eq:212}
\mu_D|_{U'} = \varphi_U^\ast (\mu_D|_U)
\end{equation}
i.e., that \((U,\mu_D|_U)\) and \((U',\mu_D|_{U'})\) are isometric with \(\varphi_U\) the isometry. Of course \(\varphi_U\) may have some fixed points (corresponding to (non-generic) closed null generators) but, for the cases of interest here, \(\varphi_U\) is not simply the identity map (even if \(U = U'\)) since, generically, the generators will not be closed. Thus open subsets of \((D,\mu_D)\) will be non-trivially isometric to other open subsets of this same space and, as we shall see from the recurrence theorem, there will be infinitely many such local isometries of \((D,\mu_D)\) due to the fact that a generic generator intersecting \textit{D} will reintersect \textit{D} in infinitely many distinct points.

\subsection{Application of the {Poincar\'{e}} Recurrence Theorem}
\label{subsec:application-poincare}
In this subsection we shall show that the {Poincar\'{e}} recurrence theorem \cite{Poincare:1899,Arnold:1978} can be applied to the flow on \textit{N} generated by the vector field \textit{X} defined in section~\ref{subsec:geometrical}. Using this theorem we shall then show that every point \(p \in N\), when mapped sufficiently far (in either direction) along the flow of \textit{X}, returns arbitrarily closely to its initial position. When combined with the isometric character of this flow (relative to the transversal metric) derived in the previous subsection, this result will lead to very stringent  restrictions upon the topological nature of the flow.

Since our spacetime \(({}^{(4)}\!V, g)\) is, by assumption, both non-compact and time-orientable, it necessarily admits a global, smooth timelike vector field \textit{V} which, without loss of generality, we may assume has been normalized to unit length (i.e., to have \(g(V,V) = -1\)). Since \textit{V} is timelike, it is necessarily transversal to the null surface \textit{N}. This follows from noting that the normalization condition, evaluated in a gaussian null coordinate chart, reduces to
\begin{equation}\label{eq:213}
-1 = g(V,V)|_N = \lbrace 2V^t V^3 + \mu_{ab} V^a V^b\rbrace|_{t=0}
\end{equation}
which clearly implies that \(V^t\) is nowhere vanishing. Expressed more invariantly this statement is equivalent to \(g (X,V)|_N \neq 0\) since, in an arbitrary agn chart adapted to \(X,\, g (X,V)|_N = V^t|_{t=0}\). Assume for definiteness that \(V^t|_{t=0} > 0\) everywhere on \textit{N} (i.e., in every agn chart adapted to \textit{X} on \textit{N}).

Following Hawking and Ellis \cite{Hawking:note}, we define a positive definite metric \(g'\) on \({}^{(4)}\!V\) by setting
\begin{equation}\label{eq:214}
g' (Y,Z) = g (Y,Z) + 2g (Y,V) g (Z,V)
\end{equation}
for any pair of vector fields \(Y,Z\) defined on \({}^{(4)}\!V\). This metric induces a Riemannian metric \({}^{(3)}\!g'\) on \textit{N} given, in an arbitrary gaussian null coordinate chart, by the expressions
\begin{equation}\label{eq:215}
\begin{split}
{}^{(3)}\!g'_{33} &= (2V^t V^t)|_{t=0}\\
{}^{(3)}\!g'_{3a} &= (2\mu_{ab} V^b V^t)|_{t=0}\\
{}^{(3)}\!g'_{ab} &= (\mu_{ab} + 2\mu_{ac} V^c \mu_{bd} V^d)|_{t=0}
\end{split}
\end{equation}
and having the natural volume element
\begin{equation}\label{eq:216}
\sqrt{\det{{}^{(3)}\!g'}} = \left.\left(2^{1/2} V^t \sqrt{\det{\mu}}\right)\right|_{t=0}.
\end{equation}
Since \(X = \frac{\partial}{\partial x^3}\) in an agn chart adapted to \textit{X}, we have
\begin{equation}\label{eq:217}
{}^{(3)}\!g' (X,X) = {}^{(3)}\!g'_{33} = (2V^t V^t)|_{t=0}
\end{equation}
as a globally defined, nowhere vanishing function on \textit{N}. Using this non-vanishing function as a conformal factor, we define a second Riemannian metric \({}^{(3)}\!g\) on \textit{N}, conformal to \({}^{(3)}\!g'\), by setting
\begin{equation}\label{eq:218}
{}^{(3)}\!g = \left.\left(\frac{1}{V^t}\right)^{2/3}\right|_{t=0} {}^{(3)}\!g'.
\end{equation}
The natural volume element of \({}^{(3)}\!g\) is thus given by
\begin{equation}\label{eq:219}
\begin{split}
\sqrt{\det{{}^{(3)}\!g}} &= \left.\left(\frac{1}{V^t}\right)\right|_{t=0} \sqrt{\det{{}^{(3)}\!g'}}\\
 &= \left.\left( 2^{1/2} \sqrt{\det{\mu}}\right)\right|_{t=0}
\end{split}
\end{equation}
Computing the divergence of \textit{X} with respect to the metric \({}^{(3)}\!g\), we find
\begin{equation}\label{eq:220}
\begin{split}
\nabla_{{}^{(3)}\!g} \cdot X &= \frac{1}{\sqrt{\det{{}^{(3)}g}}} \frac{\partial}{\partial x^i} \left(\sqrt{\det{{}^{(3)}g}} X^i\right)\\
 &= \left.\left(\frac{1}{\sqrt{\det{\mu}}} \frac{\partial}{\partial x^3} \left(\sqrt{\det{\mu}}\right)\right)\right|_{t=0}\\
 &= 0
\end{split}
\end{equation}
which vanishes by virtue of the result of Hawking and Ellis cited in the previous section (i.e., by virtue of the invariance of the transversal metric \(\mathring{\mu}_{ab}\) relative to the flow along \textit{X}). Equation (\ref{eq:220}) can be equivalently expressed as
\begin{equation}\label{eq:221}
\mathcal{L}_X \left(\sqrt{\det{{}^{(3)}\!g}}\right) = 0
\end{equation}
where \(\mathcal{L}\) signifies the Lie derivative. Thus the volume element of \({}^{(3)}\!g\) is preserved by the flow along \textit{X}.

It follows from the above that if \(\lbrace f^\lambda | \lambda \in \mathbb{R}\rbrace\) is the one-parameter family of diffeomorphisms of \textit{N} generated by \textit{X} and if \textit{D} is any measurable region of \textit{N} with volume (relative to \({}^{(3)}\!g\)) vol(\textit{D}) then \(\mathrm{vol}(f^\lambda D) = \mathrm{vol}(D)\, \forall\, \lambda \in \mathbb{R}\). Since \textit{N} is compact and \(f^\lambda\) is volume preserving, the {Poincar\'{e}} recurrence theorem may be applied and has the following consequences. Let \textit{p} be a point of \textit{N} and \textit{U} be any neighborhood of \textit{p} and, for any \(\lambda_0 \neq 0\), consider the sequence of iterates \(f^{n\lambda_0}\) (for \(n = 1, 2, \ldots\)) of \(f \equiv f^{\lambda_0}\) and the corresponding sequence of (equal volume) domains \(U, fU, f^2U, \ldots , f^nU, \ldots\). {Poincar\'{e}}'s theorem shows that there always exists an integer \(k > 0\) such that \(f^kU\) intersects \textit{U} and thus that, in any neighborhood \textit{U} of \textit{p}, there always exists a point \textit{q} which returns to \textit{U} under the sequence of mappings \(\lbrace f^n\rbrace\).

The above results together with those of the previous subsection show that any point \(p \in N\) eventually return to an arbitrarily small neighborhood of \textit{p} (after first leaving that neighborhood) when followed along the flow of \textit{X}. The reason is that since, by construction, \textit{X} has no zeros on \textit{N}, every point \(p \in N\) flows without stagnation along the integral curves of \textit{X}, first leaving sufficiently small neighborhoods of \textit{p} and then, by {Poincar\'{e}} recurrence, returning arbitrarily closely to \textit{p}.

It may happen that a point \textit{p} may actually flow back to itself, in which case the generator it lies on is closed, but for the generic points of interest here, the generators will not be closed and the flow will only take \textit{p} back arbitrarily closely to itself. 
\subsection{Implications of {Poincar\'{e}} Recurrence for the Transversal Metric}
\label{subsec:implications-poincare}
Consider a null generator \(\gamma\) of \textit{N} which passes through a point \textit{p} and let \textit{D} be a disk in \textit{N}, containing \textit{p}, which is (analytically) embedded transversally to the null generators which intersect it. If we follow \(\gamma\) starting at \textit{p} then {Poincar\'{e}} recurrence shows that we will either return to \textit{p} (in which case \(\gamma\) is closed) or else intersect \textit{D} in a sequence of points which approach \textit{p} arbitrarily closely.

The Riemannian metric \(\mu_D\) induced upon \textit{D} is analytic. Suppose for the moment that it has non-constant scalar curvature \({}^{(2)}\!R (\mu_D)\). By analyticity, \({}^{(2)}\!R (\mu_D)\) has non-zero gradient on an open dense subset of \textit{D} and thus, by the implicit function theorem, the connected level set of \({}^{(2)}\!R (\mu_D)\) passing through a point \textit{p} at which \({}^{(2)}\!R (\mu_D)\) has non-zero gradient is an analytic curve in \textit{D}, at least sufficiently near the point \textit{p}.

If \(\gamma\) is not closed, then it must reintersect \textit{D} in an infinite sequence of points \(\lbrace p_i\rbrace\) which approach \textit{p} arbitrarily closely. Furthermore, by invariance of the transversal metric along the flow, each of the \(p_i\) must lie on the same level curve of \({}^{(2)}\!R (\mu_D)\) that \textit{p} does. In fact the recurrences determined by the reintersections of \(\gamma\) with \textit{D} must densely fill the whole (connected) level set containing \textit{p}. This follows from the fact that a recurrence which carries \textit{p} to some sufficiently nearby point \(p'\) a metrical displacement \(\delta\) from \textit{p} (along the given level set of \({}^{(2)}\!R (\mu_D)\)) carries \(p'\) (again by invariance of the transversal metric along the flow) to a point \(p^{\prime\prime}\) which is displaced \(2\delta\) from \textit{p}, etc. Thus one gets recurrence by integral multiples of \(\delta\)  until eventually the recurrent points `run off the edge' of \textit{D}. Since, however, by {Poincar\'{e}} recurrence, \(\delta\) can be made arbitrarily small by choosing \(p'\) suitably from the sequence \(\lbrace p_i\rbrace\) and since one gets displacements of opposite sign by simply tracking the flow backwards, it's clear that the recurrences of \textit{p} densely fill a (connected) component of the level set of \({}^{(2)}\!R (\mu_D)\) on which \textit{p} lies. Furthermore, since each of these recurrences is induced by a local isometry of \((D, \mu_D)\), as described in section \ref{subsec:invariance}, it follows that if \textit{p} is a point at which \({}^{(2)}\!R (\mu_D)\) has non-zero gradient, then the whole connected level set of \({}^{(2)}\!R (\mu_D)\) containing \textit{p} consists of points of non-zero gradient of \({}^{(2)}\!R (\mu_D)\). Thus this entire level set (and not just a portion near \textit{p}) is an analytic curve lying in \textit{D}.

If, by contrast, \(\gamma\) is a closed generator then, by invariance of \(\mu_D\), points near \textit{p} have all of their recurrences a fixed metrical distance from \textit{p}, and thus all lie on metrical circles centered at \textit{p}. These circles are (at least generically) curves on which grad \({}^{(2)}\!R (\mu_D)\) is non-zero and hence are either densely filled by recurrences of points lying on them or else consist of points which all lie on closed generators. In either case, the interior of such a metric circle (contained in \textit{D} and centered at \textit{p}) is mapped repeatedly to itself by iterations of the isometry defined by the first recurrence. This isometry either corresponds to a `rational' rotation (in which each point advances by a rational multiple of the circumference of the circle on which it lies), in which case every point lies on a closed generator, or to an `irrational' rotation in which every metrical circle centered at \textit{p} is densely filled by the recurrences of any single point lying on it.

Thus, for the case in which \({}^{(2)}\!R (\mu_D)\) is non-constant, we find that non-closed generators densely fill smooth curves lying in \textit{D} whereas closed generators are either surrounded by other closed generators which (as a straightforward extension of the above argument shows) fill \textit{D} or else are surrounded by non-closed generators which densely fill sufficiently small circles about the given point of intersection of the closed generator with \textit{D}.

Using the connectedness and compactness of \textit{N} and the analyticity and invariance of the transversal metric it is clear that one can `analytically extend' the above argument to show that either (i) every generator of \textit{N} is closed ( a case which we have treated elsewhere), or (ii) almost every generator densely fills an analytic curve lying in any transversal embedded disk which that generator intersects. In the latter case, one may also have isolated instances of closed generators but these will, as we have seen, be surrounded by densely filling generators which are thus generic.

Consider the closure in \textit{N} of any one of these densely filling generators \(\gamma\). Let \(cl(\gamma)\) designate this subset of \textit{N}. Clearly \(cl(\gamma)\) intersects any disk \textit{D} transversal to \(\gamma\) in an analytic curve satisfying \({}^{(2)}\!R (\mu_D) = \hbox{constant}\) (since \(\gamma\) itself densely fills this curve). Locally, therefore, \(cl(\gamma)\) is obtained by translating such a transversal, analytic curve along the flow of \textit{X} and thus defines an analytic surface embedded in \textit{N}. Since \(cl(\gamma)\) is a closed subset of the compact set \textit{N}, the embedded surface defined by \(cl(\gamma)\) is thus a compact, connected embedded sub-two-manifold of \textit{N}. We want first to show that \(cl(\gamma)\) is in fact also orientable and thus, since it supports a smooth, nowhere vanishing tangent vector field (that induced by \textit{X}), that it must be diffeomorphic to a two-torus.

First, note that the value of \({}^{(2)}\!R (\mu_D)\) at a point \(p \in D \subset N\) is (by invariance of the metric \(\mu_D\) under the flow along \textit{X}) independent of the choice of disk \textit{D}. Any other transversal disk containing \textit{p} would yield the same value for the scalar curvature function at \textit{p}. Thus the transversal metric, though not really defining a metric on \textit{N}, nevertheless defines an analytic function, \({}^{(2)}\!R (\mu) : N \rightarrow R\), on \textit{N} given by setting
\begin{equation}\label{eq:236}
{}^{(2)}\!R (\mu)(p) = {}^{(2)}\!R (\mu_D)(p)
\end{equation}
for any \(p \in N\), where \textit{D} is any transversal disk containing \textit{p}. By construction, \({}^{(2)}\!R (\mu)\) is constant along the generators of \textit{N} and hence constant on the closure \(cl(\gamma)\) of any such generator. Indeed, each \(cl(\gamma)\) is just a connected component of a level set of \({}^{(2)}\!R (\mu)\).

At a generic point \(p \in N\), the differential \(d {}^{(2)}\!R (\mu)(p)\), will by analyticity, be non-zero and, by invariance of \(\mu\) along the flow of \textit{X}, this differential will be non-zero at every point along the generator \(\gamma\) which passes through \textit{p}. By continuity \(d {}^{(2)}\!R (\mu)\) will thus be non-zero everywhere on \(cl(\gamma)\) as well. Choosing a Riemannian metric \({}^{(3)}\!g'\) on \textit{N} (such as that discussed in section \ref{subsec:application-poincare}) one computes from \(d {}^{(2)}\!R (\mu)\) an associated vector field, \(\nabla {}^{(2)}\!R (\mu)\) which is everywhere non-zero and everywhere metrically perpendicular to \(cl(\gamma)\). Thus \(\nabla {}^{(2)}\!R (\mu)\) is perpendicular to \textit{X} at every point of \(cl(\gamma)\). Using the metric \({}^{(3)}\!g'\) and its associated volume 3-form one can define a `cross-product' of \textit{X} and \(\nabla {}^{(2)}\!R (\mu)\) by taking the dual of the wedge product of the corresponding one-forms and `raising the index' of the resulting one-form. This yields another smooth vector field which is tangent to \(cl(\gamma)\), nowhere vanishing and everywhere perpendicular to \textit{X}. Thus \textit{X} together with this `cross-product' vector field, define an orientation for \(cl(\gamma)\) which is thus necessarily orientable.

Therefore, any of the embedded two-manifolds, \(cl(\gamma)\), on which \(d {}^{(2)}\!R (\mu)\) is non-zero is (since compact, orientable and supporting a smooth non-vanishing vector field) necessarily a two-torus. By analyticity these are generic, since \(d {}^{(2)}\!R (\mu)\) can vanish only on isolated curves (corresponding to closed generators) or two-manifolds. The latter are necessarily tori as well since they can be shown to be orientable by a different argument.

To see this, we need to show that the compact two-manifold \(cl(\gamma)\) can be assigned a smooth, nowhere vanishing normal field. Let \textit{p} be a point in \(cl(\gamma)\) and let \textit{D} be a disk in \textit{N} (transversal to the flow of \textit{X} as usual) which contains \textit{p} as an interior point. We know that \(cl(\gamma)\) intersects \textit{D} in an analytic curve and that the recurrences of \textit{p}, followed to the future along the integral curve of \textit{X} through \textit{p}, densely fill this curve in \textit{D}. Suppose that one of these future recurrences of \textit{p} is a point \(p' \in D \cap cl(\gamma)\) which lies a metrical distance \(\delta\) (as measured along the curve \(D \cap cl(\gamma)\) with respect to the metric \(\mu_D\)) from the point \textit{p}. The point \(p'\) is uniquely determined by the point \textit{p} and the distance \(\delta\) since if, on the contrary, there were another future recurrence point \(p^{\prime\prime}\) of \textit{p}, an equal distance from \textit{p} (but on the opposing side of the curve \(D \cap cl(\gamma)\) from \(p'\)), then the same isometry which carries \textit{p} to the future to \(p'\) would carry \(p^{\prime\prime}\) to the future to \textit{p}. But this would imply that \(\gamma\) is closed which is contrary to our assumption that \(cl(\gamma)\) is a closed two-manifold densely filled by \(\gamma\).

The same isometry which uniquely carries \textit{p} to \(p'\) carries any point \(q \in D \cap cl(\gamma)\), sufficiently near to \textit{p}, to a uniquely determined point \(q' \in D \cap cl(\gamma)\) a metrical distance \(\delta\) from \textit{q} (as measured, as before, along the curve \(D \cap cl(\gamma)\) by means of the metric \(\mu_D\)). It now follows from translating \textit{D} along the flow of \textit{X} and appealing to the invariance of the transversal metric and the fact that \(\gamma\) densely fills \(cl(\gamma)\) that any point \(q \in cl(\gamma)\) lies in a transversal disk \(D_q\) which also contains a uniquely defined future recurrent point \(q'\) which lies a metrical distance \(\delta\) along \(D_q \cap cl(\gamma)\) from \textit{q} (as measured by the transversal metric).

A unique vector can now be defined at \textit{q} which is orthonormal (as measured relative to the Riemannian metric \({}^{(3)}\!g'\) defined on \textit{N}) to the embedded two-manifold \(cl(\gamma)\). To see this, choose a disk \(D_q\) containing \textit{q} (e.g., a translate along the flow of \textit{X} of the original disk \textit{D}) which intersects \(cl(\gamma)\) in an analytic arc which contains the unique future recurrent point \(q'\) a metrical distance \(\delta\) from \textit{q}. By parametrizing this arc with an orientation defined by the direction leading from \textit{q} to \(q'\) (along the segment of length \(\delta\)) we can compute a vector at \textit{q} by calculating the tangent vector at this point. This vector depends upon the choice of disk \(D_q\) but, after taking its cross product with \textit{X} (using the metric \({}^{(3)}\!g'\) as before) and normalizing to unit length, we get a uniquely defined unit normal vector to \(cl(\gamma)\) at the arbitrary point \textit{q}. That this choice varies smoothly with the choice of point \(q \in cl(\gamma)\) can be seen as follows. The parametrized arc through \textit{q} has a smoothly varying tangent. Translating this curve along the flow of \textit{X} and appealing to the invariance of the transversal metric we can generate locally (i.e., on a neighborhood of \textit{q} in \(cl(\gamma)\) a smooth tangent field to \(cl(\gamma)\) which, together with the cross product and normalization construction described above, determines a locally smooth unit normal field to \(cl(\gamma)\). However, this normal field is globally unique and thus, since smooth on a neighborhood of any point of \(cl(\gamma)\), defines a globally smooth normal direction to \(cl(\gamma)\). Thus \(cl(\gamma)\) is, as before, a compact, orientable embedded two-manifold in \textit{N} which supports a nowhere vanishing vector  field (e.g., \textit{X} or the cross product of \textit{X} with the normal field). As such it must be a torus.

Thus the level sets of \({}^{(2)}\!R (\mu)\) in \textit{N} consist of at most a finite collection of closed generators (by compactness of \textit{N} and the fact that these circles are isolated for the cases of interest here) together with a foliation of the complement of these circles by embedded two-tori. Each closed generator (if any exist) lies at the core of a family of nested tori. Each torus in the complement of the closed generators is densely filled by an integral curve of \textit{X}, i.e., by the generator \(\gamma\) whose closure \(cl(\gamma)\) defines the chosen torus. In fact, from the invariance of the transversal metric along the flow of \textit{X}, it follows that every integral curve of \textit{X} lying in \(cl(\gamma)\) is densely filling. Thus there are no fixed points (\textit{X} is nowhere zero) or periodic orbits lying in any \(cl(\gamma) \approx T^2\).

The only cases which remain to be considered are those for which \({}^{(2)}\!R (\mu_D)\) is a constant on some transversal disk \textit{D}. By analyticity it follows that \({}^{(2)}\!R (\mu)\) is necessarily constant everywhere on \textit{N}. Evidently, there are three distinct possibilities corresponding to the metric \(\mu_D\) (defined on any transversal disk) being spherical (\({}^{(2)}\!R (\mu_D) > 0\)), pseudo-spherical (\({}^{(2)}\!R (\mu_D) < 0\)), or flat (\({}^{(2)}\!R (\mu_D) = 0\)). We shall show for the first two of these cases that again the closure, \(cl(\gamma)\), of any non-closed generator \(\gamma\) is an embedded, compact two-manifold diffeomorphic to \(T^2\). For the third case, when \(\mu_D\) is flat, another possibility arises, which we shall call `ergodic', in which a generator \(\gamma\) can densely fill \textit{N} itself. That such ergodic Cauchy horizons actually occur in solutions of Einstein's equations can be seen by taking the flat Kasner solution and spatially compactifying it, with suitable identifications, to yield a vacuum spacetime defined on \(T^3 \times \mathbb{R}\) which has a Cauchy horizon \(N \approx T^3\). The most obvious identification leads to a Cauchy horizon with all generators being closed but one can exploit the spatial homogeneity of the Kasner solution to make an `irrational shift' in the coordinates of the points being identified in such a way that the null generators of the Cauchy horizon \textit{N} now densely fill \textit{N}. One can of course also do this in such a way that the generators again only densely fill two-tori instead of \(T^3\). Nevertheless, the ergodic case does exist. We shall not deal with it here but mention the conjecture that every ergodic solution is essentially equivalent to (i.e., finitely covered by) one of the ergodic flat-Kasner solutions described above.

Assume now that \({}^{(2)}\!R (\mu)\) is a non-zero constant on \textit{N}, and let \textit{p} be an arbitrary point of \textit{N}. Choose a circular transversal disk \(D_p (\delta)\) centered at \textit{p} and having radius \(\delta\) (as measured along radial geodesics of the spherical or pseudo-spherical metric \(\mu_D\)). Let \textit{p} flow to the future along \textit{X} until it first reintersects \(D_p (\delta)\) at some interior point \(p'\). By assumption \(p'\) is the \textit{first} future recurrence of \textit{p} to the interior of \(D_p (\delta)\). Let \(\delta - \varepsilon > 0\) be the radial distance from \textit{p} to \(p'\) and let \(D_{p'} (\varepsilon/2) \subset D_p (\delta)\) be a circular disk of radius \(\varepsilon/2\) centered at \(p'\). We know that there is a unique isometry, determined by the flow along \textit{X}, which carries the corresponding disk \(D_p (\varepsilon/2)\) centered at \textit{p} to \(D_{p'} (\varepsilon/2)\). This isometry is the restriction of an orientation preserving isometry of the sphere or pseudo-sphere to the subdomain defined by \(D_p (\varepsilon/2)\) and, as such, belongs to a uniquely defined one-parameter subgroup of the full (spherical or pseudo-spherical) orientation preserving isometry group.

The action of this subgroup is generated by a unique Killing field \textit{K} of the manifold \((D, \mu_D)\). From Killing's equation, \(\mathcal{L}_K \mu_D = 0\), one gets that \(\mu_D (K,K)\), the squared length of \textit{K}, is constant along the orbits of the one-parameter subgroup generated by \textit{K} (i.e., \(\mathcal{L}_K \left(\mu_D (K,K)\right) = 0\) on \textit{D}). Since \(\mu_D (K,K)\) is analytic and non-constant (since we have excluded the flat case for the present), its level sets are analytic curves which coincide with the orbits generated by \textit{K}. Let \(c_p\) be the orbit through \textit{p} generated by \textit{K}; this is just a connected component of the level set of \(\mu_D (K,K)\) determined by the value of this function at \textit{p}. What we want to show is that every future recurrence of \textit{p}, sufficiently near \textit{p}, actually returns to, and in fact, densely fills, the curve \(c_p\). This will guarantee, by arguments similar to those given above, that the closure of the orbit \(\gamma\) of  \textit{X} through \textit{p} is in fact a torus embedded in \textit{N} as before.

First note that not only is \(p'\) the first future recurrence of \textit{p} to the disk \(D_p (\delta)\) but also the first recurrence of \textit{p} to the smaller disk \(D_p (\delta - \varepsilon/2)\). Indeed, by choosing \(\eta > 0\), small enough it is clear that we can ensure that \(p'\) is the first future recurrence of \textit{p} to any disk of the type \(D_q (\delta - \varepsilon/2)\) where the distance \(d(q,p)\) from \textit{p} to \textit{q} (as measured by the metric \(\mu_D\)) is less than \(\eta\). In particular, we clearly need \(\eta < \varepsilon/2\) but let us take \(\eta\) sufficiently small so that the disk, \(D_p (\eta)\), of radius \(\eta\) centered at \textit{p}, intersects the level set of \(\mu_D (K,K)\) corresponding to the level value \(\mu_D (K,K) (p)\) only along an arc of \(c_p\) (i.e., if this level set includes disconnected components we choose \(\eta\) small enough so that \(D_p (\eta)\) excludes them). Further require (if necessary) that \(\eta < \varepsilon/4\) so that any point of the disk \(D_q (\delta - \varepsilon/2)\), for which \(d(q,p) < \eta < \varepsilon/4\), is at least a distance greater than \(\eta\) from the boundary of the original disk \(D_p (\delta)\). This ensures that the first recurrence of any point \(q \in D_p (\eta)\) to the disk \(D_q (\delta - \varepsilon/2)\) must be given by that isometry which carried \textit{p} to \(p'\) (and \(D_p (\varepsilon/2)\) to \(D_{p'} (\varepsilon/2)\)). The reason is that, if this were not the case, then the distinct isometry which first carries \textit{q} to some \(q' \in D_q (\delta - \varepsilon/2)\) would take \textit{p} to some point \(p^{\prime\prime}\) distinct from \(p'\) (since we are excluding the case of a closed generator through \textit{p}) which lies within \(D_p (\delta)\) (since \(p^{\prime\prime}\) lies within a distance \(\eta\) of \(q'\) and every point of \(D_q (\delta - \varepsilon/2)\) is at least a distance \(\eta\) from the boundary of \(D_p (\delta)\)). But this contradicts the original assumption that \(p'\) was the first future recurrence of \textit{p} to \(D_p (\delta)\).

Thus the first future recurrence of any \(q \in D_p (\eta)\) to the disk \(D_q (\delta - \varepsilon/2)\) is in fact that \(q'\) which is determined by the unique isometry which carries \(D_p (\varepsilon/2)\) to \(D_{p'} (\varepsilon/2)\) (and, of course, \textit{p} to \(p'\)).

Now, let \(q \in D_p (\eta)\) be some subsequent future recurrence of \textit{p} to \(D_p (\eta)\). We want to show that \(q \in c_p\) so suppose this is not the case. This would mean that \textit{q} and its image \(q'\) (under the isometry which carries \(D_p (\varepsilon/2)\) to \(D_{p'} (\varepsilon/2)\)) lie on some other level set of \(\mu_D (K,K)\) corresponding to a level value different from that determined by \(c_p\) (i.e., different from \(\mu_D (K,K) (p)\)). This is impossible however, since the point \(q'\) represents the first future recurrence of \textit{q} to \(D_q (\delta - \varepsilon/2)\) whereas \textit{q} is a future recurrence of \textit{p}. But the invariance of the transversal metric along the flow of \textit{X} implies the triple \((q, q', D_q (\delta - \varepsilon/2))\) must be an isometric copy of the triple \((p, p', D_p (\delta - \varepsilon/2))\) which results from simply translating the original triple along the flow until \textit{p} gets mapped to \textit{q}, etc. However, that means that \textit{p} and \textit{q} (as well as \(p'\) and \(q'\)) must both lie on the same level of \(\mu_D (K,K)\) and hence both lie on \(c_p\).

Thus all (future) recurrences of \textit{p} sufficiently near \textit{p} must lie on the analytic curve \(c_p \subset D_p (\delta)\) which contains \textit{p}. A completely analogous argument shows that the same is true for past recurrences of \textit{p}. Since these recurrences must approach \textit{p} or, in fact, any of its recurrences on \(c_p\) arbitrarily closely it is clear that, as before, the recurrences of \textit{p} densely fill the analytic curve \(c_p \subset D_p (\delta)\). Translating this curve along the flow generated by \textit{X} yields an analytic surface through \textit{p} defined locally by the foregoing constructions. Thus near any point \(p \in N\) the closure \(cl(\gamma)\), of the orbit of \textit{X} through \textit{p} is an analytically embedded two-dimensional submanifold of \textit{N}. Since \(cl(\gamma)\) is closed in \textit{N} and \textit{N} is compact, \(cl(\gamma)\) must as before be a compact submanifold of \textit{N} which supports a (smooth) nowhere vanishing vector field (e.g. \textit{X} itself). We can now use the same argument as that given above for those (isolated) manifolds having \(\nabla {}^{(2)}\!R (\mu) = 0\) to show that \(cl(\gamma)\) is orientable and hence a torus.

This argument breaks down in the case \({}^{(2)}\!R (\mu) = 0\) (i.e., when \(\mu\) is flat) but only if the isometry carrying \textit{p} to \(p'\) is a pure translation (since then and only then is \(\mu_D (K,K)\) constant on \textit{D}). The flat case still allows special cases for which \(cl(\gamma)\) is a two-torus and in those instances the arguments to follow go through equally well. But the flat case also allows more general patterns of recurrence in which \(cl(\gamma)\) is not simply a 2-manifold, but may in fact be all of \textit{N}. We shall refer to these more general cases as `ergodic' and shall not deal with them in the following. It is worth noting, however, that if an `ergodic' flow on \textit{N} generated by \textit{X} happened to admit a global transversal foliation with closed leaves (i.e., compact embedded two-manifolds everywhere transversal to the flow of \textit{X} and intersected by every orbit) then we could treat this case as well by a modification of the arguments to be given below.

Thus the picture we have developed that \textit{N} contains, at most, a finite number of closed generators and that any non-closed generator \(\gamma\) yields an embedded two-torus in \textit{N} as its closure applies to every case except the ergodic ones for which \({}^{(2)}\!R (\mu)\) is necessarily zero.

\subsection{A Connection on \textit{N} and some associated `ribbon arguments'}
\label{subsec:connection}
Let \({}^{(4)}\!Y\) and \({}^{(4)}\!Z\) be any two smooth vector fields on \(({}^{(4)}\!V, g)\) which are tangent to \textit{N} (i.e., for which \({}^{(4)}\!Y^t|_{t=0} = {}^{(4)}\!Z^t|_{t=0} = 0\) in an arbitrary gaussian null coordinate chart). Then, computing the covariant derivative \(\nabla_{{}^{(4)}\!Y} {}^{(4)}\!Z\), determined by the spacetime metric \textit{g}, observe that the resulting vector field is automatically also tangent to \textit{N} as a consequence of the invariance property of the transversal metric which was derived in Sect.~\ref{subsec:invariance} (i.e., of the result that \(\mathring{\mu}_{ab,3} = 0\)). This fact, which corresponds to the vanishing of the connection components \(\Gamma_{ij}^t|_{t=0}\) (for \(i,j = 1, 2, 3\)), in turn implies that \textit{N} is \textit{totally geodesic} (i.e., that every geodesic of \textit{g} initially tangent to \textit{N} remains in \textit{N} through its entire interval of existence).

If \textit{Y} and \textit{Z} designate the vector fields on \textit{N} induced by \({}^{(4)}\!Y\) and \({}^{(4)}\!Z\) respectively, then we can, by virtue of the above remarks, define a connection \({}^{(3)}\!\Gamma\) on \textit{N} by means of the following defining formula for covariant differentiation
\begin{equation}\label{eq:222}
{}^{(3)}\!\nabla_Y Z \equiv \left.\left(\nabla_{{}^{(4)}\!Y} {}^{(4)}\!Z\right)\right|_N.
\end{equation}
Here the right hand side symbolizes the vector field naturally induced on \textit{N} by \(\nabla_{{}^{(4)}\!Y} {}^{(4)}\!Z\). A straightforward computation in gaussian null coordinate charts (restricted to \textit{N}) shows that
\begin{equation}\label{eq:223}
\left({}^{(3)}\!\nabla_Y Z\right)^k = Y^j Z_{\hphantom{k},j}^k + {}^{(3)}\!\Gamma_{ij}^k Z^i Y^j
\end{equation}
where
\begin{equation}\label{eq:224}
{}^{(3)}\!\Gamma_{ij}^k = \Gamma_{ij}^k|_{t=0}
\end{equation}
and where \(\Gamma_{\beta\gamma}^{\alpha}\) are the Christoffel symbols of \(g_{\alpha\beta}\). The components of \({}^{(3)}\!\Gamma\) are given explicitly by
\begin{equation}\label{eq:225}
\begin{split}
{}^{(3)}\!\Gamma_{33}^3 &= -\frac{1}{2} \mathring{\varphi}_{,t}\,,\;\;\; {}^{(3)}\!\Gamma_{a3}^3 = -\frac{1}{2} \mathring{\beta}_{a,t}\\
{}^{(3)}\!\Gamma_{ab}^3 &= -\frac{1}{2} \mathring{\mu}_{ab,t}\,,\; {}^{(3)}\!\Gamma_{33}^d = 0\\
{}^{(3)}\!\Gamma_{3a}^d &= 0,\;\;\;\;\qquad {}^{(3)}\!\Gamma_{ab}^d = {}^{(2)}\!\mathring{\Gamma}_{ab}^d
\end{split}
\end{equation}
where \({}^{(3)}\!\Gamma_{ij}^k = {}^{(3)}\!\Gamma_{ji}^k\) and where the \({}^{(2)}\!\mathring{\Gamma}_{ab}^d\) are the Christoffel symbols of the invariant transversal metric \(\mathring{\mu}_{ab} (x^c)\).

A similar calculation shows that if \({}^{(4)}\!\Omega\) is a one-form on \(({}^{(4)}\!V, g)\) and \(\Omega\) its pull-back to \textit{N} then the pull-back of \(\nabla_{{}^{(4)}\!Y} {}^{(4)}\!\Omega\) is given by \({}^{(3)}\!\nabla_Y \Omega\) where, as expected,
\begin{equation}\label{eq:226}
\left({}^{(3)}\!\nabla_Y \Omega\right)_i = Y^j \Omega_{i,j} - {}^{(3)}\!\Gamma_{ij}^k Y^j \Omega_k.
\end{equation}

Now, recall the fixed vector field \textit{X} which was introduced in Sect.~\ref{subsec:geometrical}, and, for simplicity, work in agn charts adapted to \textit{X} so that \(X = \frac{\partial}{\partial x^3}\). For an arbitrary vector field \textit{Z} defined on \textit{N} we find, by a straightforward computation, that
\begin{equation}\label{eq:227}
{}^{(3)}\!\nabla_Z X = \left(\omega_X (z)\right) X
\end{equation}
where \(\omega_X\) is a one-form given, in the agn charts adapted to \textit{X}, by
\begin{equation}\label{eq:228}
\omega_X = -\frac{1}{2} \mathring{\varphi}_{,t} dx^3 - \frac{1}{2} \mathring{\beta}_{a,t} dx^a.
\end{equation}

The exterior derivative of \(\omega_X\) is readily found to be
\begin{equation}\label{eq:229}
\begin{split}
d\omega_X = &-\frac{1}{2} (\mathring{\varphi}_{,ta} - \mathring{\beta}_{a,t3}) dx^a \wedge dx^3\\
 &-\frac{1}{2} \mathring{\beta}_{a,tb} dx^b \wedge dx^a.
\end{split}
\end{equation}
However, the Einstein equation \(R_{3b} = 0\), restricted to \textit{N} and reduced through the use of \(\mathring{\mu}_{ab,3} = 0\), becomes (c.f. Eq.~(3.2) of Ref.~\cite{Moncrief:1983}):
\begin{equation}\label{eq:230}
\mathring{\varphi}_{,ta} - \mathring{\beta}_{a,t3} = 0.
\end{equation}
Thus \(d\omega_X\) reduces to
\begin{equation}\label{eq:231}
d\omega_X = -\frac{1}{2} \mathring{\beta}_{a,tb} dx^b \wedge dx^a.
\end{equation}

In subsequent sections, we shall be studying integrals of the form
\begin{equation}\label{eq:232}
\int_\gamma \omega_X = \int_\gamma \left(-\frac{1}{2} \mathring{\varphi}_{,t}\right) dx^3
\end{equation}
along segments \(\gamma\) of integral curves of \textit{X}. We shall be interested in comparing the values of these integrals for nearby integral curves. For that purpose, the following sort of \textit{ribbon argument} will prove indispensable.

Let \textit{p} and \(p'\) be any two points of \textit{N} which can be connected by a smooth curve which is everywhere transversal to the flow of \textit{X}. Let \(c : I \rightarrow N\) be such a curve defined on the interval \(I = [a,b]\) with \(c(a) = p\) and \(c(b) = p'\) and let \(\ell : I \rightarrow \mathbb{R}\) be a smooth, strictly positive function on \textit{I}. Now consider the strip or \textit{ribbon} generated by letting each point \(c(s)\) of the curve \textit{c} flow along \textit{X} through a parameter distance \(\ell(s)\) (i.e., through a lapse of \(\ell(s)\) of the natural curve parameter defined by \textit{X}). This construction gives an immersion of the ribbon
\begin{equation}\label{eq:233}
r = \left\lbrace (s,t) \in \mathbb{R}^2 | s \in I, 0 \leq t \leq \ell(s)\right\rbrace
\end{equation}
into \textit{N} which consists of connected segments of integral curves of \textit{X}. In particular, the integral curves starting and \textit{p} and \(p'\) form the \textit{edges} of the ribbon whereas the initial curve \textit{c} together with its image after flow along \textit{X} form the \textit{ends} of the ribbon.

If \(i : r \rightarrow N\) is the mapping which immerses \textit{r} in \textit{N} according to the above construction and \(i^\ast\omega_X\) and \(i^\ast d\omega_X\) are the pull-backs of \(\omega_X\) and \(d\omega_X\) to \textit{r} respectively, then one sees from Eq.~(\ref{eq:231}) and the tangency of the ribbon to the integral curves of \textit{X}, that \(i^\ast d\omega_X = 0\).

Therefore, by means of Stokes' theorem, we get
\begin{equation}\label{eq:234}
\int_{\partial r} \omega_X = \int_r d\omega_X = 0
\end{equation}
for any ribbon of the type described above. Thus if \(\gamma\) and \(\gamma'\) designate the two edges of \textit{r} (starting from \(s = a\) and \(s = b\) respectively and oriented in the direction of increasing \textit{t}) and if \(\sigma\) and \(\sigma'\) designate the two \textit{ends} of \textit{r} defined by \(\sigma = \left\lbrace (s,0) | s \in I\right\rbrace\) and \(\sigma' = \left\lbrace \left(s,\ell (s)\right) | s \in I\right\rbrace\) respectively (and oriented in the direction of increasing \textit{s}) then we get, from \(\int_{\partial r} \omega_X = 0\), that
\begin{equation}\label{eq:235}
\int_\gamma \omega_X - \int_{\gamma'} \omega_X = \int_\sigma \omega_X - \int_{\sigma'} \omega_X.
\end{equation}

Equation (\ref{eq:235}) will give us a means of comparing \(\int_\gamma \omega_X\) with \(\int_{\gamma'} \omega_X\) provided we can estimate the contributions to \(\int_{\partial r} \omega_X\) coming from the ends of the ribbon. As a simple example, suppose (as we did in Ref.~\cite{Moncrief:1983}) that every integral curve of \textit{X} is closed and choose \textit{r} and \(i : r \rightarrow N\) so that the image of \textit{r} in \textit{N} consists of a ribbon of simply closed curves. In this case, the end contributions cancel and we get that \(\int_\gamma \omega_X = \int_{\gamma'} \omega_X\). This result played a key role in the arguments of Ref.~\cite{Moncrief:1983}. 
\section{Elementary Regions and their Analytic Foliations}
\label{sec:elementary}
In the sections to follow we shall define a `candidate' vector field \textit{K} on \textit{N} by rescaling \textit{X} appropriately, prove its analyticity and eventually show that \textit{K} propagates into the enveloping spacetime as an analytic Killing vector field. If, for some reason, we knew a priori that \textit{N} admitted a global, analytic foliation with closed leaves that are everywhere transverse to the flow of \textit{X} then we could proceed with this analysis much as we did for the (higher dimensional) stationary black holes of Ref.~\cite{Moncrief:2008}, working `globally' on \textit{N} by directly exploiting the special structure provided by its `pre-existing' analytic foliation. Here however no such analytic foliation has been presumed to exist and indeed the very possibility of global, closed, transversal leaves might be excluded for purely topological reasons.\footnote{For example, even for the case of closed generators of \textit{N} the integral curves of \textit{X} might well be the fibers of a non-trivial \(\mathbf{S}^1\)-bundle as is indeed the case for the Taub-NUT family of spacetimes.} On this account we shall decompose \textit{N}, as needed, into a finite collection of \textit{elementary regions} that will each be shown to admit an analytic, transversal foliation and carry out the aforementioned analysis first on the individual elementary regions, much as we did for the case of closed generators in Ref.~\cite{Moncrief:1983}. Finally, after verifying the consistency of these constructions on overlapping domains of definitions, we shall assemble the resulting components and ultimately arrive at a globally defined, analytic `candidate' vector field \textit{K} on \textit{N}.

Consider any one of the analytically embedded 2-tori discussed in Sect.~(\ref{subsec:implications-poincare}) that is realized as the closure, \(cl(\gamma)\), of a (non-closed but densely-torus-filling) generator \(\gamma\). This torus supports the flow of a nowhere-vanishing, analytic vector field, namely that induced from \textit{X} which, by construction, is tangential to the chosen embedded torus.

Thanks to a theorem due to \textit{M}. Kontsevic (of which the proof is sketched below in the Appendix) one knows that such a torus always admits an analytic foliation with closed leaves that are everywhere transverse to the flow generated by \textit{X}. We now wish to `thicken' such an embedded torus to obtain an embedded 3-manifold diffeomorphic to \(\mathcal{A} \times \mathbf{S}^1\) (where \(\mathcal{A}\) is an open annulus), consisting entirely of generators of \textit{N}, and to show that this thickened torus will itself admit an analytic foliation (with leaves each diffeomorphic to \(\mathcal{A}\)) that is everywhere transverse to the flow generated by \textit{X}. Such a thickened torus, together with its analytic, transverse foliation, will be the first of two types of \textit{elementary regions} that we shall define.

The second type of elementary region will only be needed to cover a `tubular neighborhood in \textit{N}' of any particular \textit{closed} generator \(\gamma\) that might, exceptionally, occur. In this case we shall `thicken' \(\gamma\) to a solid torus diffeomrophic to \(\mathcal{D} \times \mathbf{S}^1\) (where \(\mathcal{D}\) is an open disk), consisting entirely of generators sufficiently close to \(\gamma\), and show that such a solid torus admits an analytic, transversal of foliation with leaves diffeomorphic to \(\mathcal{D}\). For the case of a non-ergodic flow (as defined in Sect.~(\ref{subsec:implications-poincare})) every generator of \textit{N} is either closed or densely fills an embedded 2-torus. By the compactness of \textit{N} such a null hypersurface can clearly be covered by a finite collection of such elementary regions with those of the  second type only needed in the presence of closed generators.

To construct such elementary regions we shall need an analytic, Riemannian metric on \textit{N}. To define such a metric we slightly modify the argument in Sect.~(\ref{subsec:application-poincare}) by now insisting that the (normalized, timelike) vector field \textit{V}, which is transverse to \textit{N} in \(({}^{(4)}\!V, g)\), be itself analytic. Since the timelike condition is an \textit{open} one and since the normalization of such an analytic vector field will not disturb its analyticity there is no loss of generality involved in assuming that the induced Riemannian metric \({}^{(3)}\!g\) is in fact analytic on \textit{N}. Recall that the metric so defined (via Eqs.~(\ref{eq:213})--(\ref{eq:218}) in fact satisfies
\begin{equation}\label{eq:401}
\mathcal{L}_X \left(\sqrt{\det{{}^{(3)}\!g}}\right) = 0
\end{equation}
on \textit{N}.

As discussed in the Appendix one constructs an analytic, transversal foliation with closed leaves for any one of such embedded 2-tori by showing that it always admits an analytic, \textit{closed} one-form \(\lambda\) with integral periods that, moreover, satisfies \(\lambda (X) > 0\). Since any such \(\lambda\) is locally expressible as \(\lambda = d\omega\) for some analytic function \(\omega\), the level sets of \(\omega\) define the leaves of the foliation. Thus \(\omega\) provides an analytic coordinate function that is constant of the leaves so-defined. The closure of these leaves and their transversality to \textit{X} is ensured by the integrality of the periods of \(\lambda\) and by the condition that \(\lambda (X) > 0\) everywhere on the torus. Any two such coordinate functions, \(\omega\) and \(\omega'\), will of course only differ by a constant on their overlapping domains of definition.

We now `thicken' the chosen 2-torus by flowing along the normal geodesics of the metric \({}^{(3)}\!g\) on \textit{N}, much as we would  in constructing a \textit{gaussian-normal} neighborhood of the given torus. By restricting the range of the (normal geodesic) flow parameter suitably one can ensure that the resulting thickened torus is diffeomorphic to \(\mathcal{A} \times \mathbf{S}^1\), where \(\mathcal{A}\) is an open annulus corresponding to a thickened leaf of the original torus, and consists entirely of integral curves of \textit{X}. By continuity, if this thickening is sufficiently restricted the annular leaves of the foliated 3-manifold will be globally transverse to \textit{X}.

We  now extend the domain of definition of the analytic, coordinate function \(\omega\) by requiring it to be everywhere constant on any one of the thickened leaves. Choosing complementary, analytic coordinates\ \(\lbrace x^a\rbrace = \lbrace x^1, x^2\rbrace\) on one of these annular leaves and holding these fixed along the flow generated by \textit{X} while setting \(x^3 = \omega\) one gets a convenient adapted coordinate chart for the thickened torus \(\approx \mathcal{A} \times \mathbf{S}^1\). Any two such coordinate systems \(\lbrace x^a, x^3\rbrace\) and \(\lbrace x^{a'}, x^{3'}\rbrace\) will be related, on their overlapping domains of definition by a transformation of the form
\begin{equation}\label{eq:402}
\begin{split}
x^{3'} &= x^3 + \hbox{ constant}\\
x^{a'} &= f^a (x^1, x^2)
\end{split}
\end{equation}
where the \(\lbrace f^a\rbrace\) define an analytic diffeomorphism of the annulus \(\mathcal{A}\). Thus this first type of elementary region consists of a thickened 2-torus foliated, on the one hand, by the (non-closed) integral curves of \textit{X} and, on the other, by annuli transverse to the flow of \textit{X}.

The second type of elementary region results from thickening a closed generator \(\gamma\) to get a solid torus with \(\gamma\) at its core. To construct this choose an analytic, `angle' coordinate \(x^3\) to label the points of the chosen generator \(\gamma\). At each point \textit{p} of \(\gamma\) we have a corresponding, orthogonal 2-plane in the tangent space, \(T_p N\), defined by the metric \({}^{(3)}\!g\) (i.e., the orthogonal complement to the tangent vector to \(\gamma\) at \textit{p}). By flowing along the geodesics of \({}^{(3)}\!g\) in \textit{N} we may thus `thicken' each such point \(p \in \gamma\) to a disk \(\mathcal{D}_p\) which, by construction, is orthogonal to \(\gamma\) at \textit{p}. By restricting the geodesic flow parameter suitably (in its dependence upon \textit{p} and the orthogonal direction to \(\gamma\) at \textit{p}) we may ensure that the \(\gamma\) so thickened is diffeomorphic to \(\mathcal{D} \times \mathbf{S}^1\), consists entirely of integral curves of \textit{X} and is such that each thickened leaf, \(\mathcal{D}_p\), is transverse to the flow generated by \textit{X}.

By defining an analytic coordinate \(x^3\) on \(\mathcal{D} \times \mathbf{S}^1\) by holding the chosen angular coordinate for \(\gamma\) constant on each leaf and by choosing complementary, analytic coordinates \(\lbrace x^a\rbrace = \lbrace x^1, x^2\rbrace\) for any one of the transversal disks and holding these constant along the flow of \textit{X} we generate an adapted analytic coordinate system for this second type of elementary region. Any two such coordinate systems, \(\lbrace x^a, x^3\rbrace\) and \(\lbrace x^{a'}, x^{3'}\rbrace\), will be related by a transformation of the form
\begin{equation}\label{eq:403}
\begin{split}
x^{3'} &= x^3 + \hbox{ constant}\\
x^{a'} &= g^a (x^1, x^2)
\end{split}
\end{equation}
on their overlapping domains of definition where now the \(\lbrace g^a\rbrace\) define an analytic diffeomorphism of the disk.

In the following sections it will be convenient to let the symbol \(\mathcal{H}\) designate an arbitrary elementary region of either of the two types. By the compactness of \textit{N} it is clear that we can cover \textit{N} by a finite collection of such elementary regions. 
\section{Nondegeneracy and geodesic incompleteness}
\label{sec:nondegeneracy}
In this section we shall show, using a ribbon argument, that each null geodesic generator of \textit{N} is either complete in both directions (the `degenerate' case) or else that each generator is incomplete in one direction (the non-degenerate case).  More precisely, we shall prove that if any single generator \(\gamma\) is incomplete in a particular direction (say that defined by \textit{X}) then every other generator of the (connected) hypersurface \textit{N} is necessarily incomplete in the same direction.  It will then follow that if any generator  is complete in a particular direction, then all must be since otherwise one could derive a contradiction from the first result.  We shall see later that, in the non-degenerate case, the generators which are all incomplete in one direction (say that of \textit{X}) are however all complete in the opposite direction (that of \(-X\)).

As usual we work in adapted charts for an arbitrary fundamental region \(\mathcal{H} \subset N\).  For the calculations to follow however, it is convenient to work with charts induced from adapted charts on the covering space \(\hat{\mathcal{H}}\approx\Sigma\times\mathbb{R}\) of \(\mathcal{H}\)~\footnote{Where either \(\Sigma \approx \mathcal{A}\) or \(\Sigma \approx \mathcal{D}\) depending upon the type of the elementary region \(\mathcal{H}\).} for which the \(\lbrace x^a\mid a = 1, 2\rbrace\) are constant along any given generator and the range of the `angle' coordinate \( x^3\) is unwrapped  from say \(\lbrack\mathring{x}^3, \mathring{x}^3 + s^\ast)\), where \(s^\ast\) is the `recurrence time' for \(x^3\) on \(\mathcal{H}\), to cover the interval \((-\infty, \infty)\).  Projected back to \(\mathcal{H}\) these induce families of charts \(\lbrace x^3, x^a\rbrace, \lbrace x^{3'}, x^{a'}\rbrace\), etc. related, on their regions of overlap, by analytic transformations of the form
\begin{equation}\label{eq:501}
\begin{split}
x^{3'} &= x^3 + \text{constant}\\
x^{a'} &= f^a(x^1, x^2).
\end{split}
\end{equation}
By working on the covering space we simplify the notation by keeping the \(\lbrace x^a\rbrace\) constant and letting \(x^3\) range continuously over \((-\infty, \infty)\) in following a given generator as it repeatedly sweeps through the leaves of the chosen foliation of \(\mathcal{H}\).  However, one should keep in mind that this is just an artifice to represent calculations carried out on the elementary region \(\mathcal{H}\) in a simplified notation since the compactness of the closure, \(cl(\mathcal{H})\), of \(\mathcal{H}\) in \textit{N} will play a key role in the arguments to follow.

Consider a null generator of \(\mathcal{H}\) developed from `initial' conditions specified at a point \(p\in \mathcal{H}\) having coordinates \(\lbrace x^3 (p) = \mathring{x}^3, x^a(p) = \mathring{x}^a\rbrace\). The affine parametrization of this generator is determined by solving the geodesic equations which, for the class of curves in question, effectively reduce to
\begin{equation}\label{eq:502}
\begin{split}
&\frac{d^2x^3}{d\eta^{2}} - \frac{\mathring{\varphi}_{,t}}{2}  (x^3, x^a) \left(\frac{dx^3}{d\eta}\right)^2 = 0 \\
&x^a(\lambda) = \mathring{x}^a = \,\,\text{constant}
\end{split}
\end{equation}
where \(\eta\) is an affine parameter. To complete the specification of initial conditions one needs, of course, to give an initial velocity \(\frac{dx^3}{d\eta}\mid_{\mathring{\eta}}\) (taking \(\frac{dx^a}{d\eta}\mid_{\mathring{\eta}} = 0\)).

Solving the first order equation
\begin{equation}\label{eq:503}
\frac{dv}{d\eta} = \frac{\mathring{\varphi}_{,t}}{2} \,\, v^2
\end{equation}
for \(v := \frac{dx^3}{d\eta}\) to get an integral formula for \textit{v} and then integrating \(\frac{d\eta}{dx^3} = \frac{1}{v}\) with respect to \(x^3\) one derives an expression for the affine length of a segment of this null geodesic defined on the interval \([\mathring{x}^3, x^3]\):
\begin{equation}\label{eq:504}
\begin{split}
&\eta (x^3,\mathring{x}^a) - \mathring{\eta} (\mathring{x}^3, \mathring{x}^a)\\
&=\frac{1}{(\frac{dx^3}{d\eta})}\bigm|_{\mathring{\eta} (\mathring{x}^{3},\mathring{x}^{a})} \int\limits^{x^{3}}_{\mathring{x}^{3}} d\rho \quad \text{exp} [-\int\limits^\rho_{\mathring{x}^{3}} d\xi (\frac{\mathring{\varphi}_{,t}}{2} (\xi,\mathring{x}^a)] .
\end{split}
\end{equation}
Thus incompleteness of this generator, in the direction of \(X = \frac{\partial}{\partial x^3}\), would correspond to the existence of the limit
\begin{equation}\label{eq:505}
\begin{split}
&\lim\limits_{x^{3}\to\infty} \int\limits^{x^{3}}_{\mathring{x}^{3}} d\rho \quad\text{exp} [-\int\limits^\rho_{\mathring{x}^{3}} d\xi (\frac{\mathring{\varphi}_{,t}}{2} (\xi , \mathring{x}^a))]\\
&\quad =(\frac{dx^{3}}{d\eta})\bigm|_{\mathring{\eta} (\mathring{x}^{3},\mathring{x}^a)} (\eta (\infty, \mathring{x}^a) - \mathring{\eta} (\mathring{x}^3, \mathring{x}^a)) < \infty
\end{split}
\end{equation}
whereas completeness (in this direction)  would correspond to the divergence of this limit. Recalling Equation (\ref{eq:232}), note that the integral of the one-form \(\omega_X\) along the segment \(\gamma\) defined above is given by
\begin{equation}\label{eq:506}
\int\limits_\gamma \omega_X = \int\limits^{x^{3}} _{\mathring{x}^{3}} (-\frac{1}{2} \mathring{\varphi},_{t} (\xi , \mathring{x}^a))d\xi
\end{equation}
which thus provides an invariant representation of the basic integral arising in the above formulas.

Suppose that the generator `beginning' at \(p\in \mathcal{H}\) is incomplete in the direction of \textit{X}. We want to establish convergence of the corresponding integral for any other generator of \(\mathcal{H}\).  Since incompleteness is an asymptotic issue (the relevant integrals being automatically finite on any compact domain of integration) there is no essential loss of generality in comparing only those generators that `start' in the slice defined by \textit{p}.  Thus we want to consider generators `beginning' at points \textit{q}, having \(x^3 (q) = \mathring{x}^3\), and establish their incompleteness by using a suitable ribbon argument.  Furthermore, to have a `canonical' way of defining our comparison ribbons it will be convenient to localize the calculations somewhat by first looking only at generators sufficiently near to the `reference' generator. Thus, given a point \textit{p} in the initial slice defined by \(x^3(p) = \mathring{x}^3\), we consider only those points \textit{q} lying in this slice which, additionally, lie within a \textit{closed} geodesic ball (relative to the invariant transversal metric \(\mu\) induced on this slice) centered at \textit{p} and contained within a normal neighborhood of this point.  Any such \textit{q} can be connected to \textit{p} by a unique geodesic lying within this geodesic ball and such points can be conveniently labeled by normal coordinates defined at \textit{p} (i.e., the points of a corresponding, closed ball in the tangent space to the slice at \textit{p}).

The unique geodesic connecting \textit{q} to \textit{p} provides a canonical `starting end' to our comparison ribbon for geodesics emanating from points \textit{p} and \textit{q} (in the direction of \textit{X}) and, from invariance of the transversal metric along the flow of \textit{X}, we get an isometric image of this connecting geodesic induced on any subsequent slice traversed along the flow.

Let \(\gamma\) be the segment of the null generator beginning at \textit{p} and defined on the interval \([\mathring{x}^3, x^3]\), for some \(x^3 > \mathring{x}^3\), and let \(\gamma'\) be a corresponding segment of the generator beginning at \textit{q} and defined on the same interval. From the argument given in Section \ref{subsec:connection} it follows that
\begin{equation}\label{eq:507}
\int\limits_\gamma \omega_X - \int\limits_{\gamma'} \omega_X = \int\limits_\sigma \omega_X  - \int\limits_{\sigma'} \omega_X
\end{equation}
where \(\sigma\) is the geodesic end defined in the starting slice and $\sigma'$ its isometric image at the ending slice.

For fixed \textit{p} the integral \(\int_\sigma \omega_X\) varies continuously with \textit{q} as \textit{q} ranges over a \textit{compact} set (the closed geodesic ball centered at \textit{p} described above) and thus is bounded for all \textit{q} in this ball. Furthermore the integral \(\int_{\sigma'} \omega_X\) varies continuously with \textit{q} and \(x^3\) but, as \(x^3\) increases, the image of \textit{p} under the flow ranges only over (some subset of) the \textit{compact} set given by the closure of \(\mathcal{H}\) in \textit{N} whereas the image of \textit{q} remains always a fixed geodesic distance from the image of \textit{p} in the corresponding slice. Since the product of the closure of \(\mathcal{H}\) with this (closed) ball is compact the continuously varying integral \(\int_{\sigma'} \omega_X\) (regarded as a function of \textit{q} and \(x^3\) for fixed \textit{p}) is necessarily bounded no matter how large the ``unwrapped'' coordinate \(x^3\) is allowed to become.

It follows from the forgoing that for any fixed \textit{p} and \textit{q} as above, there exists a \textit{bounded}, continuous (in fact analytic) real-valued function \(\delta_{p,q}(x^3)\) such that
\begin{equation}\label{eq:508}
\int\limits_{\gamma'} \omega_X = \int\limits_\gamma \omega_X + \delta_{p,q}(x^3)
\end{equation}
for arbitrary \(x^3 > \mathring{x}^3\).  But this implies that
\begin{equation}\label{eq:509}
\begin{split}
&\int\limits_{\mathring{x}^{3}}^{x^{3}} d\rho \,\, \text{exp} [-\int\limits_{\mathring{x}^{3}}^\rho \frac{\mathring{\varphi}_{,t}}{2} (\xi, \mathring{x}^a(q))d\xi] \\
&= \int\limits_{\mathring{x}^{3}}^{x^{3}} d\rho \,\, \text{exp} [-\int\limits_{\mathring{x}^{3}}^\rho \frac{\mathring{\varphi}_{,t}}{2} (\xi, \mathring{x}^a(p))d\xi + \delta_{p,q} (\rho)] \\
&= \int\limits_{\mathring{x}^{3}}^{x^{3}} d\rho \,\, \text{exp}[\delta_{p,q}(\rho)] \,\,\text{exp} [-\int\limits_{\mathring{x}^{3}}^\rho \frac{\mathring{\varphi}_{,t}}{2} (\xi, \mathring{x}^a(p))d\xi]  .
\end{split}
\end{equation}
From the boundedness of \(\delta_{p,q}\)
\begin{equation}\label{eq:510}
- \infty < b_1 \leq \delta_{p,q}(\rho)\leq b_2 < \infty,
\forall\rho\in [\mathring{x}^3, \infty)
\end{equation}
it follows that
\begin{equation}\label{eq:511}
\begin{split}
&e^{b_{1}} \int\limits_{\mathring{x}^{3}}^{x^{3}} d\rho \,\, \text{exp}[-\int\limits_{\mathring{x}^{3}}^\rho \frac{\mathring{\varphi}_{ ,t}}{2} (\xi, \mathring{x}^a(p))d\xi] \\
&\leq \int\limits_{\mathring{x}^{3}}^{x^{3}} d\rho \,\,\text{exp}[-\int\limits_{\mathring{x}^{3}}^\rho \frac{\mathring{\varphi}_{ ,t}}{2} (\xi, \mathring{x}^a(q))d\xi] \\
&\leq e^{b_{2}} \int\limits_{\mathring{x}^{3}}^{x^{3}} d\rho \,\, \text{exp}[-\int\limits_{\mathring{x}^{3}}^\rho \frac{\mathring{\varphi} _{,t}}{2} (\xi, \mathring{x}^a(p))d\xi]
\end{split}
\end{equation}
\(\forall x^3\in [\mathring{x}^3 ,\infty)\). But this implies that if the limit
\begin{equation}\label{eq:512}
\lim\limits_{x^{3}\to\infty} \int^{x^{3}}_{\mathring{x}^{3}} d\rho \,\, \text{exp}[-\int^\rho_{\mathring{x}^{3}} \frac{\mathring{\varphi} _{,t}}{2} (\xi, \mathring{x}^a(p))d\xi]
\end{equation}
exists, then so must the limit of the monotonically increasing function \(\int^{x^{3}}_{\mathring{x}^{3}} d\rho \,\, \text{exp}[-\int^\rho_{\mathring{x}^{3}} \frac{\mathring{\varphi}_{ ,t}}{2} (\xi, \mathring{x}^a(q))d\xi]\) exist as \(x^3\to\infty\). Conversely, if the affine length of \(\gamma\) diverges, then so must that of \(\gamma'\) by virtue of the forgoing bounds.

So far we have only considered those null generators starting within a geodesic ball centered at a point \textit{p} in the initial slice.  But from the compactness and connectedness of \textit{N} it's clear that any of its null generators can be thus compared to the original `reference generator' through a finite collection of such \textit{ribbon arguments} and thus all of them shown either to be incomplete in the direction \textit{X} or else to be complete in this direction. Clearly the same argument can be applied in the opposite direction (i.e., that of $-X$) with a corresponding conclusion. However, as we shall see later, the non-degenerate case will always be characterized by generators that are all incomplete in one direction but complete in the opposite direction, whereas the degenerate case will be characterized by generators that are complete in both directions.  
\section{A candidate vector field in the non-degenerate case}
\label{sec:candidate-vector}

In this section, we focus on the non-degenerate case and, if necessary, change the sign of $X$ so that it points in a direction of incompleteness for the null generators of \textit{N}. We now define a vector field \textit{K} on \textit{N}, also tangent to the generators of this hypersurface, by setting \(K = uX\) where \textit{u} is a positive real-valued function on \textit{N} chosen so that, for any point \(p\in N\), the null generator determined by the initial conditions \((p, K(p) = u(p) X(p))\) has a fixed (i.e., independent of \textit{p}) future affine length given by \(\frac{2}{k}\) where \textit{k} is a constant \(> 0\).  At the moment there is no preferred normalization for \textit{k} so we choose its value arbitrarily.

From Equation (\ref{eq:505}) upon putting \((\eta (\infty, \mathring{x}^a) - \mathring{\eta} (\mathring{x}^3, \mathring{x}^a)) = \frac{2}{k}\), we see that \(u(x^3, x^a)\) is necessarily expressible, in an arbitrary `unwrapped' elementary region \(\hat{\mathcal{H}}\) for \textit{N}, by
\begin{equation}\label{eq:601}
u(x^3,x^a) = \frac{k}{2} \int\limits^\infty_{x^{3}} d\rho\; \text{exp}\left[-\int\limits^{\rho}_{x^{3}} \frac{\mathring{\varphi},_{t}}{2} (\xi, x^a)d\xi\right].
\end{equation}
By the results of the previous section, the needed integral converges for every generator and clearly \(u > 0\) on \(\hat{\mathcal{H}}\).  What is not clear however, in view of the limiting procedure needed to define the outer integral over a semi-infinite domain, is whether \textit{u} is in fact analytic and we shall need to prove that it is.  We shall do this below by showing that a sequence \(\{ u_i : \mathcal{H}\to\mathbf{R}^+\mid i = 1,2,\ldots\}\) of analytic `approximations' to \textit{u} defined by
\begin{equation}\label{eq:602}
u_i(x^3,x^a) = \frac{k}{2} \int\limits^{x^{3}+is^{*}}_{x^{3}} d\rho ~ \text{exp}\left[-\int\limits^\rho_{x^{3}} \frac{\mathring{\varphi},_{t}}{2} (\xi, x^a)d\xi\right],
\end{equation}
where \(s^*\) is the recurrence time introduced in Section \ref{sec:nondegeneracy}, does indeed have an analytic limit as \(i\to\infty\).

For the moment however, let us assume that we know that \textit{K} is analytic and introduce new agn coordinates \(\{ x^{3'}, x^{a'},t'\}\) which are adapted to \textit{K} rather than to \(X = \frac{\partial}{\partial x^{3}}\). Thus we seek a transformation of the form \(\{x^{3'} = h(x^3, x^a), x^{a'} = x^a\}\) which yields \(K = \frac{\partial}{\partial x^{3'}}\). A straightforward calculation shows that \textit{h} must satisfy
\begin{equation}\label{eq:603}
\frac{\partial h(x^{3},x^{a})}{\partial x^{3}} = \frac{1}{u(x^{3},x^{a})} = \left\{\frac{1}{\frac{k}{2} \int^\infty_{x^3} d\rho ~ \exp[-\int^\rho_{x^{3}} d\xi \frac{\mathring{\varphi}_{,t}}{2}(\xi, x^{a})]}\right\}
\end{equation}
which, since the denominator is analytic by assumption and non-vanishing, yields an analytic \textit{h} upon integration.

As was shown in Sect. IIIA of Reference \cite{Moncrief:1983}, a transformation of the above type connects the primed and unprimed metric functions \(\mathring{\varphi}_{,t}\) and \(\mathring{\varphi}'_{,t'}\) via
\begin{equation}\label{eq:604}
2 \frac{\partial}{\partial x^3}\left(\frac{\partial h}{\partial x^3}\right) + \frac{\partial h}{\partial x^3} \mathring{\varphi}_{,t} = \left(\frac{\partial h}{\partial x^3}\right)^2 \, \mathring{\varphi}'_{,t'} .
\end{equation}
Computing \(\frac{\partial^2h}{\partial x^{3~2}}\) from Equation (\ref{eq:603}) above and substituting this and
\(\frac{\partial h}{\partial x^3}\) into the above formula one finds that the transformed metric has
\begin{equation}\label{eq:605}
\mathring{\varphi}'_{,t'} = k = \text{constant}
\end{equation}
throughout any agn chart adapted to \textit{K}.  This argument is somewhat the reverse of that given in Reference \cite{Moncrief:1983}, for the case of closed generators, wherein we set \(\mathring{\varphi}'_{, t'} = k\) and solved Equation (\ref{eq:604}) for \(\frac{\partial h}{\partial x^3}\) and then \textit{h}.

In the new charts one still has \(\mathring{\varphi}' = \mathring{\beta}'_\alpha = 0\) since these hold in any agn coordinate system and, upon repeating the argument of Section \ref{subsec:invariance} above, with \textit{K} in place of \textit{X}, we obtain
\(\mathring{\mu}'_{a'b',3'} = 0\) as well.  Now evaluating the Einstein equation \(R_{3b} = 0$ at $t = t' = 0\) and using the foregoing, together with the new result that \(\mathring{\varphi}'_{,t'} = k\) in the primed charts, one finds that \(\mathring{\beta}'_{b',t',3'} = 0\).

Deleting primes to simplify the notation, we thus find that in agn charts adapted to \textit{K}, the metric functions obey
\begin{equation}\label{eq:606}
\mathring{\varphi} = \mathring{\beta}_a = \mathring{\mu}_{ab,3} = 0,\,\, \mathring{\varphi}_{,t} = \text{constant} \neq 0, (\mathring{\beta}_{a,t})_{,3} = 0.
\end{equation}
These are the main results we shall need for the inductive argument of Section~\ref{sec:existence-killing} to prove that there is a spacetime Killing field \textit{Y} such that \(Y\mid_N = K\).

Referring to Equation (\ref{eq:504}) and evaluating the integrals in the new charts in which \(\mathring{\varphi}_{,t} = k = \text{constant} > 0\) one sees easily that though the null generators are all incomplete towards the `future' they are in fact all complete towards the `past' (where here future and past designate simply the directions of \textit{K} and \(-K\) respectively).  It may seem strange at first glance to say that any generator could have a fixed future affine length \((=\frac{2}{k})\) no matter where one starts along it, but the point is that this length is here always being computed from the geodesic initial conditions \((p, K(p))\).  If one starts with say \((q, K(q))\) and later reaches a point \textit{p} on the same generator, then the tangent to the (affinely parametrized) geodesic emanating from \textit{q} will not agree with \(K(p)\) but will instead equal \(c K(p)\) for some constant \(c > 1\),  Only upon `restarting' the generator with the initial conditions \((p, K(p))\) will it be found to have the same future affine length that it had when started instead from \((q, K(q))\).  Indeed, if the tangent to an affinely parametrized geodesic did not increase relative to \textit{K} then the generator could never be incomplete on a compact manifold \textit{N} where the integral curves of a vector field \textit{K} are always complete.

Let us now return to the question of the analyticity of the `scale factor' \(u(x^3,x^a)\).  First note that, upon combining Equations (\ref{eq:603}), (\ref{eq:604}) and (\ref{eq:605}), \textit{u} satisfies the linear equation with analytic coefficients
\begin{equation}\label{eq:607}
\frac{\partial u}{\partial x^3} - \frac{\mathring{\varphi}_{,t}}{2} \,\,  u = - \frac{k}{2}
\end{equation}
provided one takes, as initial condition specified at some \(\mathring{x}^3\),
\begin{equation}\label{eq:608}
u(\mathring{x}^3, x^a) = \frac{k}{2} \int\limits^\infty_{\mathring{x}^3} d\rho \,\, \text{exp}[-\int\limits^\rho_{\mathring{x}^3} \frac{\mathring{\varphi}_{,t}}{2} (\xi, x^a)d\xi ].
\end{equation}
More precisely, using an appropriate integrating factor for Equation (\ref{eq:607}), namely \(\text{exp}[-\int^{x^{3}}_{\mathring{x}^3} d\xi \frac{\mathring{\varphi}_{,t}}{2} (\xi, x^a)]\), one easily shows that the solution to Equation \ref{eq:607}) determined by the initial condition (\ref{eq:608}) is given by Equation (\ref{eq:603}).  But Equation (\ref{eq:607}) can be viewed as a (linear, analytic) partial differential equation to which the Cauchy Kowalewski theorem applies \cite{John:1991} and guarantees the analyticity of the solution on domains corresponding (because of linearity) to those of the coefficients (in this case \(\mathring{\varphi}_{,t} (x^3,x^a))\) \textit{provided} that the initial condition \(u(\mathring{x}^3, x^a)\) is analytic with respect to the \(\{ x^a\}\).  In other words, our problem reduces to that of proving that Equation (\ref{eq:608}) for fixed \(\mathring{x}^3\), defines an analytic function of the \(\{x^a\}\).  Thus we only need to show that the sequence of `approximations'
\begin{equation}\label{eq:609}
\begin{split}
&u_i(\mathring{x}^3, x^a) := \frac{k}{2} \int\limits^{\mathring{x}^{3} + is^{*}}_{\mathring{x}^{3}} d\rho \,\, \text{exp} [-\int\limits^\rho_{\mathring{x}^{3}} \frac{\mathring{\varphi}_{,t}}{2} (\xi, x^a)d\xi],\\
&i = 1, 2, \ldots
\end{split}
\end{equation}
converges to an analytic function of the \(\{x^a\}\) for fixed \(\mathring{x}^3\).

However, a (pointwise) convergent sequence of analytic functions could easily converge to a limit which is not even continuous much less analytic.  On the other hand, the set of continuous functions on a compact manifold forms a Banach space with respect to the \(C^0\) norm (uniform convergence) so that one could hope at least to establish the continuity of the limit by showing that the sequence \(\{u_i (\mathring{x}^3, x^a)\}\) is Cauchy with respect to this norm.

A much stronger conclusion is possible however, if one first complexifies the slices \(x^3 = \text{constant}\) of an arbitrary elementary region \(\mathcal{H} \subset N\) (which are each diffeomorphic to a manifold \(\Sigma\) of the type defined previously) and extends the analytic metric functions defined on \textit{N} to holomorphic functions defined on this complex `thickening' of \(\mathcal{H}\) in the \(\{ x^a\}\) directions which extend continuously to the boundary of its closure.  The space of holomorphic functions on such a complex manifold (with boundary) forms a Banach space with respect to the \(C^0\) norm so that the limit of any Cauchy sequence of holomorphic functions (which extend continuously to the boundary) will in fact be holomorphic and not merely continuous \cite{Treves:2006,Ohsawa:note}.  In the following section, we shall define a certain complex `thickening' of \textit{N} with respect to all of its dimensions (a so-called `Grauert tube') but then, in view of the discussion in the preceding paragraph, restrict the integration variable \(x^3\) defined on an aribtrary elementary region \(\mathcal{H}\) to real values so that, in effect, only the leaves of the foliation of \(\mathcal{H} \approx \Sigma\times \mathbf{S}^1\) are thickened.

Let us temporarily remain within the real analytic setting to sketch out the basic idea of the argument to be given later in the holomorphic setting.  This detour, though it cannot yield more than the continuity of \(u(\mathring{x}^3, x^a)\) in the \(\{ x^a\}\) variables, will be easier to understand at a first pass and will require only straightforward modification for its adaptation to the holomorphic setting.

For any point \textit{p} in the slice determined by \(x^3(p) = \mathring{x}^3\) the monotonically increasing, convergent sequence of real numbers
\begin{equation}\label{eq:610}
\begin{split}
u_i(\mathring{x}^3, x^a(p)) &= \frac{k}{2} \int\limits^{\mathring{x}^{3} + is^{*}}_{\mathring{x}^{3}} d\rho  \,\, \text{exp} [-\int\limits^\rho_{\mathring{x}^{3}} d\xi \,\, \frac{\mathring{\varphi}_{,t}}{2} (\xi, x^a(p))] \\
&i = 1,2, \ldots
\end{split}
\end{equation}
is clearly a Cauchy sequence which converges to \(u(\mathring{x}^3, x^a(p))\).  Thus for any \(\varepsilon' > 0\) there exists a positive integer \textit{Q} such that
\begin{equation}\label{eq:611}
\mid u_m(\mathring{x}^3, x^a(p)) - u_\ell (\mathring{x}^3, x^a(p))\mid < \varepsilon' \quad  \forall \,\, m, \ell > Q.
\end{equation}
Now consider an arbitrary point \textit{q} in the initial slice (i.e., having \(x^3(q) = \mathring{x}^3\)) that lies within a closed geodesic ball in this slice which is centered at \textit{p} (i.e., a ball of the type used in the ribbon argument of the previous section).  By the ribbon arguments given in this last section, one easily finds that
\begin{equation}\label{eq:612}
\begin{split}
&\mid u_m(\mathring{x}^3, x^a(q)) - u_\ell (\mathring{x}^3, x^a(q))\mid \\
&= \biggm|\frac{k}{2}\int\limits^{\mathring{x}^{3} + ms^{*}}_{\mathring{x}^{3}+\ell s^{*}} \,\,d\rho \,\, \text{exp}[-\int\limits^\rho_{\mathring{x}^3} \,\, d\xi \, \frac{\mathring{\varphi}_{,t}}{2} (\xi, x^a(q))]\biggm| \\
&= \biggm|\frac{k}{2} \int\limits^{\mathring{x}^{3} + ms^{*}}_{\mathring{x}^{3}+\ell s^{*}} \,\,d\rho \,\, \text{exp}[\delta_{p,q}(\rho)] \text{exp}[-\int\limits^\rho_{\mathring{x}^3} \,\, d\xi \, \frac{\mathring{\varphi}_{,t}}{2} (\xi, x^a(p))]\biggm|  \\
&\leq e^{b_{2}} \biggm|\frac{k}{2} \int\limits^{\mathring{x}^{3} + ms^{*}}_{\mathring{x}^{3}+\ell s^{*}} \,\,d\rho \,\, \text{exp}[-\int\limits^\rho_{\mathring{x}^3} \,\, \frac{\mathring{\varphi}_{,t}}{2} (\xi, x^a(p))d\xi]\biggm| \\
&= e^{b_{2}} \bigm| u_m(\mathring{x}^3, x^a(p)) - u_\ell (\mathring{x}^3, x^a(p))\bigm|
\end{split}
\end{equation}
for all \textit{q} in this ball where \(b_2\) is a constant that depends upon \textit{p} and the radius of the chosen ball.  Thus for any \(\varepsilon > 0\) we get by choosing \(\varepsilon'  = e^{-b_{2}}\varepsilon\) in Equation (\ref{eq:611}), that
\begin{equation}\label{eq:613}
\bigm| u_m (\mathring{x}^3, x^a(q)) - u_\ell (\mathring{x}^3, x^a(q))\bigm| \,\, < \varepsilon \quad \forall \, m, \ell > Q
\end{equation}
and for all \textit{q} in the compact set defined by the chosen (closed) geodesic ball.  Thus the sequence of (real-valued) continuous functions \(\{ u_m(\mathring{x}^3, x^a(q))\mid m = 1,2,\ldots\}\) defined on this ball is a Cauchy sequence relative to the \(C^0\)-norm and hence its limit \(u(\mathring{x}^3, x^a(q))\) is necessarily continuous.  By covering the initial slice by a collection of overlapping such balls, we deduce that \(u(\mathring{x}^3, x^a(q))\) is globally continuous on the initial slice.  
\section{Analyticity of the candidate vector field}
\label{sec:analyticity}
Recall from Section~\ref{subsec:application-poincare} that one can define a Riemannian metric \({}^{(3)}\!g\) on the horizon manifold \textit{N} that satisfies \(\mathcal{L}_X \sqrt{\det{{}^{(3)}\!g}} = 0\). From the discussion in Section~\ref{sec:elementary} it is clear that this metric can always be chosen to be analytic so that in fact \((N, {}^{(3)}\!g)\) is a compact, analytic, Riemannian 3-manifold.

There is a canonical way of complexifying a compact, analytic Riemannian manifold such as \((N, {}^{(3)}\!g)\)through the introduction of its so-called Grauert tubes \cite{Burns:2003}.  One identifies \textit{N} with the zero section of its tangent bundle \(TN\) and defines a map \(\ell :TN\to\mathbf{R}\) such that \(\ell (v)\) is the length of the tangent vector \(v\in TN\) relative to the Riemannian metric \({}^{(3)}\!g\).  Then, for sufficiently small \(s > 0\), the manifold (`Grauert tube' of thickness \textit{s})
\begin{equation}\label{eq:702}
T^sN = \{ v\in TN \mid \ell (v) < s\}
\end{equation}
can be shown to carry a complex structure for which holomorphic coordinates \(\{ z^i\}\) can be defined in terms of analytic coordinates \(\{ x^i\}\) for \textit{N} by setting \(z^k = x^k + iy^k\) where \(y = y^k \frac{\partial}{\partial x^{k}}\) represents a vector in \(TN\).  Analytic  transformations between overlapping charts for \textit{N} extend to holomorphic transformations between  corresponding charts for \(T^sN\) provided that, as we have assumed, \textit{N} is compact and \textit{s} is sufficiently small.  For non-compact manifolds such a holomorphic thickening need not exist for any \textit{s}, no matter how small, and further restrictions upon the manifold are in general needed in order to define its Grauert tubes. When defined, Grauert tubes have an anti-holomorphic involution \(\sigma: T^sN\to T^sN\) given by \(v\mapsto -v\).

It will be convenient to define an auxiliary, analytic Riemannian metric, \(g_{\mathcal{H}}\), on each elementary region of interest \(\mathcal{H}\) by writing on \(\hat{\mathcal{H}} \approx \Sigma \times \mathbb{R}\),
\begin{equation}\label{eq:722}
\begin{split}
g_{\mathcal{H}} &= (g_{\mathcal{H}})_{ij} dx^i \otimes dx^j\\
&= dx^3 \otimes dx^3 + \mu_{ab} (x^1, x^2) dx^a \otimes dx^b
\end{split}
\end{equation}
and then, as before, identifying the slice at \(x^3\) with that at \(x^3 + s^\ast\) via the aforementioned analytic isometry of \((\Sigma, \mu)\). This metric is adapted to the chosen slicing of \(\mathcal{H}\) in that each \(x^3 = \mathrm{constant}\) slice is a totally geodesic submanifold of \((\mathcal{H}, g_{\mathcal{H}})\) and furthermore the integral curves of \(X = \frac{\partial}{\partial x^3}\), which is evidently a Killing field of \(g_{\mathcal{H}}\), coincide with the geodesics of \((\mathcal{H}, g_{\mathcal{H}})\) normal to the \(x^3 = \mathrm{constant}\) slices.

From the special properties of the metric \(g_{\mathcal{H}}\) and its geodesics, it is easy to see that if \(\{ x^a\mid a= 1, 2\}\) are normal coordinates for \((\Sigma, \mu)\) centered at a point \(q\in\Sigma\) (with, therefore, \(x^a(q) = 0\)) then, holding these constant along the flow of \textit{X} and, complementing them with the function \(x^3\), we get normal coordinates \(\{ x^i\} = \{ (x^a, x^3)\mid a = 1, 2\}\) defined on a tubular domain in \(\mathcal{H}\) centered on the orbit of \textit{X} through \textit{q}.  By shifting \(x^3\) by an additive constant, one can of course arrange that the origin of these normal coordinates for this tubular domain lies at any chosen point along the orbit through \textit{q}.  It follows from the aforementioned property of Grauert tubes that the functions
\begin{equation}\label{eq:703}
\{ z^k\} = \{ (z^k = x^k + iy^k) \mid (y^3)^2 + \mu_{ab} (x^1, x^2) y^a y^b < s\}
\end{equation}
will provide, for \textit{s} sufficiently small, holomorphic coordinates on a corresponding complex thickening of \(\mathcal{H}\) which we shall denote by \(T^s\mathcal{H}\).

In the application to follow, as already mentioned in the previous section, we shall set \(y^3 = 0\) and thus focus our attention on `thickenings' of \(\mathcal{H}\) of the restricted form \(T^s\Sigma\times\mathbf{S}^1\) which are foliated by curves of the type
\begin{equation}\label{eq:704}
\begin{split}
z^a(\lambda) &= x^a(\lambda) + iy^a(\lambda) = \mathring{x}^a + i\mathring{y}^a \\
&= \text{constant}, \\
z^3(\lambda) &= \mathring{x}^3 + \lambda, \quad y^3(\lambda) = 0,
\end{split}
\end{equation}
with
\begin{equation}\label{eq:705}
\mu_{ab} (\mathring{x}^1, \mathring{x}^2) \mathring{y}^a\mathring{y}^b < s.
\end{equation}
The closure \(\overline{T^s\Sigma\times\mathbf{S}^1}\approx \overline{T^s\Sigma}\times\mathbf{S}^1\), of this manifold results from attaching a boundary to \(T^s\Sigma\times\mathbf{S}^1\) characterized locally by \(\mu_{ab} (x^1,x^2)y^ay^b = s\) at all points \((x^1, x^2) \in \overline{\Sigma}\) and will also play a role in the considerations to follow.

Analytic tensor fields defined on \textit{N} can always, in view of its compactness, be lifted to define holomorphic fields on thickenings of the type \(T^s\mathcal{H}\) which, furthermore, extend continuously to the boundary of \(\overline{T^s\mathcal{H}}\) provided \(s > 0\) is taken to be sufficiently small.  The needed limitation on the size of \textit{s} arises from considering the radii of convergence of the local series representations of these fields on the original analytic manifold \textit{N} but, since it is compact, a finite collection of such representations suffices to define the field globally on \textit{N} and hence a choice of \(s > 0\) is always possible so that a given field on \textit{N} extends holomorphically to \(T^sN\).  Upon restricting such a field to the manifold \(T^s\Sigma\times\mathbf{S}^1\), as defined by setting \(y^3 = 0\), one obtains a corresponding field that is holomorphic with respect to the \(\{ z^a \mid a = 1, 2\}\), real analytic with respect to \(x^3\) and which extends continuously to the boundary of \(\overline{T^s\Sigma\times\mathbf{S}^1} \approx \overline{T^s\Sigma}\times\mathbf{S}^1\). From our point of view, the important thing is that such fields form a Banach space with respect to the \(C^0\) norm and hence a Cauchy sequence with respect to this norm will necessarily converge to a holomorphic field with respect to the \(\{ z^a\}\).

To carry out ribbon arguments on the associated complex thickenings over \(\mathcal{H}\), we need to lift the one form \(\omega_X\), defined in Section~\ref{subsec:connection}, to its holomorphic correspondent \(^{(c)}\omega_X\),
\begin{equation}\label{eq:706}
\begin{split}
^{(c)}\omega_X = &-\frac{1}{2}\,\,^{(c)}\mathring{\varphi}_{,t} (z^1,\ldots, z^3)(dx^3 + idy^3) \\
& - \frac{1}{2}\,\,^{(c)}\mathring{\beta}_{a,t} (z^1,\ldots, z^3)(dx^a + idy^a)
\end{split}
\end{equation}
with
\begin{equation}\label{eq:707}
\begin{split}
^{(c)}\mathring{\varphi}_{,t} (x^1,\ldots, x^3) & = \mathring{\varphi}_{,t} (x^1,\ldots, x^3) \\
^{(c)}\mathring{\beta}_{a,t} (x^1,\ldots, x^3) &= \mathring{\beta}_{a,t} (x^1,\ldots, x^3),
\end{split}
\end{equation}
defined on a suitable \(T^s\mathcal{H}\), where the components \(^{(c)}\mathring{\varphi}_{,t} (z^1,\ldots, z^3)\) and \(^{(c)} \mathring{\beta}_{a,t} (z^1,\ldots, z^3)\) each satisfy the Cauchy-Riemann equations (ensuring their holomorphicity)
\begin{equation}\label{eq:708}
\begin{split}
&\frac{\partial}{\partial\overline{z}^{k}}\,\, ^{(c)}\mathring{\varphi}_{,t} (z^1, \ldots, z^3) \\
&= \frac{1}{2} (\frac{\partial}{\partial x^{k}} + i\frac{\partial}{\partial y^{k}})^{(c)}\varphi_{,t} (x^1,\ldots, x^3, y^1,\ldots, y^3) \\
&= 0\qquad k = 1, \ldots, 3
\end{split}
\end{equation}
and similarly for \(\frac{\partial}{\partial\overline{z}^{k}} ^{(c)}\mathring{\beta}_{a.t}(z_1,\ldots, z^3)\).  As a holomorphic one-form \(^{(c)}\omega_X\) has exterior derivative
\begin{equation}\label{eq:709}
\begin{split}
d^{(c)}\omega_X &= - \frac{1}{2} [\frac{\partial}{\partial z^{a}}\,\, ^{(c)}\mathring{\varphi}_{,t} - \frac{\partial^{(c)}}{\partial z^{3}}\mathring{\beta}_{a,t}] \\
&\cdot (dx^a + idy^a)\wedge (dx^3 + idy^3) \\
&- \frac{1}{2} \frac{\partial ^{(c)}\mathring{\beta}_{a,t}}{\partial z^{b}}  (dx^b + idy^b)\wedge (dx^a+ idy^a)
\end{split}
\end{equation}
which, in view of the complexified Einstein equation (c.f., Equation (3.2) of Reference \cite{Moncrief:1983}),
\begin{equation}\label{eq:710}
\frac{\partial^{(c)}\mathring{\varphi}_{,t}}{\partial z^{a}(z)} - \frac{\partial^{(c)}\beta_{a,t}(z)}{\partial z^{3}} = 0,
\end{equation}
reduces to
\begin{equation}\label{eq:711}
d^{(c)} \omega_X = -\frac{1}{2} \frac{\partial^{(c)}\mathring{\beta}_{a,t}}{\partial z^{b}} dz^b \wedge dz^a .
\end{equation}

For our purposes, it is convenient to regard Equation (\ref{eq:711}) as an equation for an ordinary, complex-valued, one form defined on a real analytic manifold of 6 dimensions with local coordinates
\begin{equation}\label{eq:712}
\{ w^\mu \mid \mu = 1, \ldots, 6\} = \{ x^1, \ldots, x^3, y^1, \ldots , y^3\}
\end{equation}
and with \(^{(c)}\omega_X\) decomposed into its real and imaginary parts as
\begin{equation}\label{eq:713}
^{(c)}\omega_X = \{ (^{(c)}\omega_X^{(r)}(w))_\mu + i(^{(c)}\omega^{(i)}_X (w))_\mu\} dw^\mu .
\end{equation}
By appealing to the Cauchy-Riemann equations satisfied by the components, it is easy to show that the left hand side of Equation (\ref{eq:711}) is equal to the `ordinary' exterior derivative of \(^{(c)}\omega_X\), as rewritten above, with respect to its 6 real coordinates \(\{ w^\mu\} = \{ x^1,\ldots, x^3, y^1, \ldots, y^3\}\).  The right hand side of this equation can of course be expressed in the analogous way --- as a complex-valued two-form in the same real variables.

We are now in a position to apply Stokes's theorem much as in the previous section, the only real difference being that now the one-form in question, \(^{(c)}\omega_X\) is complex and its domain of definition is a 6-real-dimensional Grauert tube defined over \(\mathcal{H}\).  We shall want to compare integrals of \(^{(c)}\omega_X\) over different curves of the type (\ref{eq:704}) extending from some `initial' slice having \(x^3 = \text{constant}\) to another such `final' slice.  For convenience, let us always take one such curve (which will provide a reference `edge' for our comparison ribbon) to lie in the real section (i.e., to have \(y^a(\lambda) = y^3(\lambda) = 0\)) and choose normal coordinates for \((\Sigma,\mu)\) so that points on this reference curve have \(x^a(\lambda) = 0\).  As in the previous section, we restrict the domain of definition of these normal coordinates to a geodesic ball relative to the metric \(\mu\).  Let \textit{p} be the starting point of this curve so that, in the chosen coordinates \(\{ x^a(p) = y^a(p) = y^3(p) = 0, x^3(p) = \mathring{x}^3\}\).

Now suppose that \(q\in T^s\mathcal{H}\) is a point lying in the domain of the corresponding (complex) chart and having \(x^3(q) = \mathring{x}^3, y^3(q) = 0\), \(\mu_{ab} (x^1(q), x^2(q))y^a(q)y^b(q) < s\) where \(\{ x^1(q), x^2(q)\}\) represents a point in the aforementioned geodesic ball centered at \textit{p}.  We want a canonical way of connecting \textit{q} to \textit{p} within the initial slice \(x^3 = \mathring{x}^3\) and, for this purpose, first connect \textit{q} to its projection in the real section with the `straight line'
\begin{equation}\label{eq:714}
\begin{split}
&x^i(\sigma) = x^i(q) = \text{constant} \\
&y^a(\sigma) = -\sigma y^a(q), \sigma\in [-1,0] \\
&y^3(\sigma) = 0.
\end{split}
\end{equation}
We complete the connection to \textit{p} along the geodesic
\begin{equation}\label{eq:715}
\begin{split}
&x^a(\sigma) = (1 - \sigma ) x^a(q), \,\, \sigma\in [0,1] \\
&x^3(\sigma) = x^3 (p) = x^3(q) = \mathring{x}^3 \\
&y^a(\sigma) = y^3(\sigma) = 0.
\end{split}
\end{equation}
This broken curve provides the starting end  (at \(x^3 = \mathring{x}^3\)) for our comparison ribbon.  We complete the specification of such a ribbon by letting each point on the starting end defined above, flow along the corresponding curve of the form (\ref{eq:704}) (i.e., holding \(x^a\) and \(y^a\) constant, \(y^3 = 0\) and letting \(x^3 = \mathring{x}^3 + \lambda\) vary until the final slice is reached).  It is easy to see, from the special form of the right hand side of
Equation (\ref{eq:711}) that the corresponding two-form pulled back to such a ribbon vanishes identically and thus that Stokes's theorem applies to integrals of \(^{(c)}\omega_X\) over its edges and ends in essentially the same way that we discussed in Section V for ribbons confined to the real section.  In other words, the integral of \(^{(c)}\omega_X\) over the edge beginning at \textit{q}, differs from that over the reference edge beginning at \textit{p} only by the (difference of) the integrals over the ribbon ends lying in the `initial' and `final' slices.

For our purposes, the contribution from the starting end, connecting \textit{q} and \textit{p}, will be fixed whereas the contribution from the `final' end (connecting the images of \textit{q} and \textit{p} induced on the final slice) will vary continuously but only over a compact set (determined by the endpoint of the edge through \textit{q} which necessarily lies in \(\overline{T^{s}\Sigma\times\mathbf{S}^{1}}\)).  Thus, if as before, we designate the edges through \textit{p} and \textit{q} by \(\gamma\) and \(\gamma'\) respectively and the initial and final ribbon ends by $\sigma$ and $\sigma'$ respectively, then we obtain, as in the real setting,
\begin{equation}\label{eq:716}
\begin{split}
&\int\limits_{\gamma'}\,\, ^{(c)}\omega_X = \int\limits_\gamma\,\,^{(c)}\omega_X - (\int\limits_\sigma \,\,^{(c)}\omega_X - \int\limits_{\sigma'} \,\,^{(c)}\omega_X) \\
&\qquad = \int\limits_\gamma \,\,  ^{(c)}\omega_X + \,\,^{(c)}\delta_{p,q} (x^3)  \\
\end{split}
\end{equation}
with
\begin{equation}\label{eq:717}
\mid^{(c)}\delta_{p,q} (\rho) \mid \leq b < \infty \quad \forall \, \rho \in [\mathring{x}^3, \infty).
\end{equation}
The integrals of course are now in general complex in value but, given the bound above, we are in a position to apply ribbon arguments to the complex setting in complete parallel to those we gave in the real setting at the end of the last section.  The arguments needed are so similar to those given previously that we shall only sketch their highlights below.

For any \textit{q} within the domain characterized above, we define a sequence
\begin{equation}\label{eq:718}
\begin{split}
&^{(c)}u_i (\mathring{x}^3, z^a(q)) \\
&= \frac{k}{2} \int\limits_{\mathring{x}^{3}}^{\mathring{x}^3 + is^*} d\rho \,\, \text{exp} [-\int\limits^\rho_{\mathring{x}^{3}} d\xi \frac{\mathring{\varphi}_{,t}}{2} (\xi, z^a(q))]
\end{split}
\end{equation}
of holomorphic extensions (to \(T^s\Sigma\times\mathbf{S}^1\)) of the approximations given earlier in Equation (\ref{eq:609}) for the normalizing function \textit{u}.  Using ribbon arguments to compare the integrals \(\int_{\gamma'}\,\, ^{(c)}\omega_X\) with those for the reference curves \(\int_\gamma \,\, ^{(c)}\omega_X\) we derive, as before, a bound of the form
\begin{equation}\label{eq:719}
\begin{split}
&\mid\,\, ^{(c)}u_m(\mathring{x}^3,  z^a(q)) - \,\,^{(c)}u_\ell (\mathring{x}^3, z^a(q))\mid \\
&\leq e^b\mid \,\,^{(c)} u_m (\mathring{x}^3, z^a(p)) - \,\,^{(c)} u_\ell (\mathring{x}^3, z^a(p))\mid \\
&= e^b \mid u_m (\mathring{x}^3, x^a(p)) - u_\ell (\mathring{x}^3, x^a(p))\mid \\
&\qquad \forall \,\, \ell, m\geq 0,
\end{split}
\end{equation}
where, in the final equality, we have exploited the fact that \(^{(c)} u_m(\mathring{x}^3, z^a(p)) = u_m (\mathring{x}^3, x^a(p))\) by virtue of our choice that the point \textit{p} always lies in the real section.

As before, it follows immediately that for any \(\varepsilon > 0\) there exists an integer \(Q > 0\) such that
\begin{equation}\label{eq:720}
\mid\,\,^{(c)}u_m(\mathring{x}^3, z^a(q)) -\,\, ^{(c)}u_\ell (\mathring{x}^3, z^a (q))\mid < \varepsilon \quad \forall \,\, m, \ell > Q
\end{equation}
and thus that the sequence \(\{^{(c)}u_m(\mathring{x}^3, z^a(q)) \mid m = 1, 2, \ldots \}\)
is Cauchy with respect to the \(C^0\) norm.  Thus the sequence of approximations converges to a holomorphic limit on the domain indicated.  Repeating this argument for a (finite) collection of such domains sufficient to cover \(\overline{T^s\Sigma}\) we conclude that
\begin{equation}\label{eq:721}
^{(c)}u(\mathring{x}^3, z^a) = \frac{k}{2} \int^\infty_{\mathring{x}^{3}} d\rho \,\, \text{exp}[-\int^\rho_{\mathring{x}^{3}} d\xi \frac{\mathring{\varphi}_{,t}}{2}(\xi, z^a)]
\end{equation}
is a well-defined holomorphic function on \(T^s\Sigma\) (which extends continuously to its boundary) and that, by construction, this function reduces to the real-valued function \(u(\mathring{x}^3, x^a)\) defined in the previous section.  The latter is therefore necessarily a real-valued analytic function on \(\Sigma\) which is the result we were required to prove.

The analytic functions thus defined on tubular neighborhoods of arbitrary null generators of \textit{N} necessarily coincide on overlapping domains of definition. This follows from the fact that each such \textit{u} was uniquely determined by the geometrical requirement that it `renormalize' the corresponding generators to all have the same, fixed future affine length \(2/k\). We may thus regard \textit{u} as a globally defined analytic function on \textit{N} and thus arrive at a globally defined, analytic, candidate vector field \(K := uX\).  
\section{Existence of a Killing Symmetry}
\label{sec:existence-killing}
We have shown that there exists a non-vanishing, analytic vector field \textit{K} on \textit{N}, tangent to the null generators of \textit{N} such that, in any gaussian null coordinate chart adapted to \textit{K} (i.e., for which \textit{K} has the local expression \(K = \left.\frac{\partial}{\partial x^3}\right|_{t=0}\)), the metric functions \(\lbrace\varphi, \beta_a, \mu_{ab}\rbrace\) of that chart obey
\begin{equation}\label{eq:301}
\begin{split}
\mathring{\varphi} = \mathring{\beta}_a &= \mathring{\mu}_{ab,3} = 0,\\
\mathring{\varphi}_{,t} = k &= \hbox{ constant } \neq 0,\\
\left(\mathring{\beta}_{a,t}\right)_{,3} &= 0.
\end{split}
\end{equation}
We shall show momentarily that \((\mathring{\mu}_{ab,t})_{,3}\) also vanishes and thus that all the metric functions and their first time derivatives are independent of \(x^3\) on the initial surface \(t = 0\) (signified as before by an overhead `nought'). In the following, we shall prove inductively that all the higher time derivatives of the metric functions are independent of \(x^3\) at \(t = 0\) and thus that the corresponding \textit{analytic}, Lorentzian metric,
\begin{equation}\label{eq:302}
\begin{split}
g &= dt \otimes dx^3 + dx^3 \otimes dt + \varphi dx^3 \otimes dx^3\\
 &+ \beta_a dx^a \otimes dx^3 + \beta_a dx^3 \otimes dx^a + \mu_{ab} dx^a \otimes dx^b,
\end{split}
\end{equation}
has \(\frac{\partial}{\partial x^3}\) as a (locally defined) Killing field throughout the gaussian null coordinate chart considered. Finally, we shall show that the collection of locally defined Killing fields, obtained by covering a neighborhood of \textit{N} by adapted gaussian null (agn) coordinate charts and applying the construction mentioned above, fit together naturally to yield a spacetime Killing field \textit{Y} which is analytic and globally defined on a full neighborhood of \textit{N} and which, when restricted to \textit{N}, coincides with the vector field \textit{K}.

Some of the results to be derived are purely local consequences of Einstein's equations expressed in an agn coordinate chart (such as, e.g., the observation that \(\mathring{\varphi}_{,t} = k\) implies \((\mathring{\beta}_{a,t})_{,3} = 0\)). Others, however, require a more global argument and thus demand that we consider the transformations between overlapping, agn charts which cover a neighborhood of \textit{N} in \({}^{(4)}\!V\). For example, by considering the Einstein equations \(R_{ab} = 0\) restricted to \(t = 0\) and reduced through the use of \(\mathring{\varphi}_{,t} = k = \hbox{ constant }, \mathring{\mu}_{ab,3} = 0\) and \((\mathring{\beta}_{a,t})_{,3} = 0\) one can derive (as in the derivation of Eq.~(3.26) of Ref.~\cite{Moncrief:1983}) the local equation for \(\mathring{\mu}_{ab,t}\) given by
\begin{equation}\label{eq:303}
0 = -(\mathring{\mu}_{ab,t})_{,33} + \frac{k}{2} (\mathring{\mu}_{ab,t})_{,3}.
\end{equation}
Roughly speaking, we want to integrate this equation along the null generators of \textit{N} and show, as in Ref.~\cite{Moncrief:1983}, that it implies that \((\mathring{\mu}_{ab,t})_{,3} = 0\). Now, however, since the null generators are no longer assumed to be closed curves, this argument requires a more invariant treatment than was necessary in Ref.~\cite{Moncrief:1983}.

First, let \(\lbrace x^\mu\rbrace = \lbrace t, x^3, x^a\rbrace\) and \(\lbrace x^{\mu'}\rbrace = \lbrace t', x^{3'}, x^{a'}\rbrace\) be any two gaussian null coordinate charts which are adapted to \textit{K} (i.e., for which \(K = \frac{\partial}{\partial x^3}|_{t=0}\) and \(K = \frac{\partial}{\partial x^{3'}}|_{t' =0}\) on the appropriate domains of definition of the given charts). It is not difficult to see that, if the two charts overlap on some region of \textit{N}, then within that region the coordinates must be related by transformations of the form
\begin{equation}\label{eq:304}
\begin{split}
x^{3'} &= x^3 + h(x^a)\\
x^{a'} &= x^{a'} (x^b)
\end{split}
\end{equation}
where \(t = t' = 0\) since we have restricted the charts to \textit{N}. Here \textit{h} is an analytic function of the coordinates \(\lbrace x^a\rbrace\) labeling the null generators of \textit{N} and \(x^{a'} (x^b)\) is a local analytic diffeomorphism allowing relabeling of those generators within the region of overlap of the charts.

We let \(\lbrace\varphi, \beta_a, \mu_{ab}\rbrace\) designate the agn metric functions of the unprimed chart,
\begin{equation}\label{eq:305}
\begin{split}
g &= g_{\mu\nu} dx^\mu \otimes dx^\nu\\
 &= dt \otimes dx^3 + dx^3 \otimes dt + \varphi dx^3 \otimes dx^3\\
 &+ \beta_a dx^a \otimes dx^3 + \beta_a dx^3 \otimes dx^a + \mu_{ab} dx^a \otimes dx^b,
\end{split}
\end{equation}
and \(\lbrace\varphi', \beta'_a, \mu'_{ab}\rbrace\) designate the corresponding functions in the primed chart.

In the region of \({}^{(4)}\!V\) in which the charts overlap, we have of course,
\begin{equation}\label{eq:306}
g_{\mu'\nu'} = \frac{\partial x^\alpha}{\partial x^{\mu'}} \frac{\partial x^\beta}{\partial x^{\nu'}} g_{\alpha\beta}
\end{equation}
and, because of the gaussian null metric form,
\begin{equation}\label{eq:307}
\begin{split}
g_{t't'} &= 0 = \frac{\partial x^\alpha}{\partial t'} \frac{\partial x^\beta}{\partial x^{t'}} g_{\alpha\beta}\\
g_{t'3'} &= 1 = \frac{\partial x^\alpha}{\partial t'} \frac{\partial x^\beta}{\partial x^{3'}} g_{\alpha\beta}\\
g_{t'a'} &= 0 = \frac{\partial x^\alpha}{\partial t'} \frac{\partial x^\beta}{\partial x^{a'}} g_{\alpha\beta}
\end{split}
\end{equation}
By virtue of the form of (\ref{eq:304}), we also have, of course, that \(\frac{\partial}{\partial x^3}|_{t=0} = \frac{\partial}{\partial x^{3'}}|_{t'=0}\) on the region of overlap (since both charts were adapted to \textit{K} by assumption).

Writing out Eqs.~(\ref{eq:307}) in more detail, using the explicit form of \(g_{\alpha\beta}\), restricting the result to the surface \(t' = t = 0\) and making use of the transformations (\ref{eq:304}) which hold on that surface, one readily derives that
\begin{equation}\label{eq:308}
\begin{split}
\left.\left(\frac{\partial t}{\partial t'}\right)\right|_{t'=0} &= 1,\\
\left.\left(\frac{\partial x^a}{\partial t'}\right)\right|_{t'=0} &= \left.(\mu^{ab} h_{,b})\right|_{t=0}\\
\left.\left(\frac{\partial x^3}{\partial t'}\right)\right|_{t'=0} &= \left.\left(-\frac{1}{2} \mu^{ab} h_{,a} h_{,b}\right)\right|_{t=0}.
\end{split}
\end{equation}
Differentiating these equations with respect to \(x^{3'}\) and using the fact that \(\mathring{\mu}_{ab,3} = 0\) one finds that
\begin{equation}\label{eq:309}
\left.\left(\frac{\partial^2 x^\alpha}{\partial x^{3'} \partial t'}\right)\right|_{t'=0} = 0.
\end{equation}
The remaining metric transformation equations (\ref{eq:307}), restricted to the initial surface, yield the covariance relation
\begin{equation}\label{eq:310}
\left.\vphantom{\frac{1}{2}}\mu_{a'b'}\right|_{t'=0} = \left.\left(\frac{\partial x^c}{\partial x^{a'}} \frac{\partial x^d}{\partial x^{b'}} \mu_{cd}\right)\right|_{t=0}
\end{equation}
as well as reproducing equations such as \(\varphi'|_{t'=0} = 0\), and \(\beta_{a'}|_{t'=0} = 0\) which are common to all gaussian null coordinate systems.

Now take the first \(t'\) derivative of the transformation Eqs.~(\ref{eq:306}), restrict the results to the surface \(t' = t = 0\) and make use of Eqs.~(\ref{eq:301}) to derive expressions for
\begin{equation}\label{eq:311}
\left.\vphantom{\frac{1}{2}}\lbrace\varphi'_{,t'}, \beta_{a',t,}, \mu_{a'b,t'}\rbrace\right|_{t'=0}
\end{equation}
in terms of unprimed quantities. Differentiating the resulting equations with respect to \(x^{3'}\) leads to the covariance relation
\begin{equation}\label{eq:312}
\left.\vphantom{\frac{1}{2}}\mu_{a'b',t'x^{3'}} \right|_{t'=0} = \left.\left(\frac{\partial x^c}{\partial x^{a'}} \frac{\partial x^d}{\partial x^{b'}} \mu_{cd,t3}\right)\right|_{t=0}
\end{equation}
as well as reproducing known results such as \(\beta_{a',t'3'}|_{t'=0} = 0\) which hold in all agn coordinate systems.

Now in any agn coordinate chart restricted to \textit{N}, we have the locally defined analytic functions
\begin{equation}\label{eq:313}
\begin{split}
D &\equiv \frac{\det{(\mathring{h}_{ab})}}{\det{(\mathring{\mu}_{ab})}}\\
T &\equiv \mathring{\mu}^{ab} \mathring{h}_{ab}
\end{split}
\end{equation}
where \(\mathring{h}_{ab} \equiv \mathring{\mu}_{ab,t3}\) and where \(\det{(~)}\) signifies determinant. From the covariance relations (\ref{eq:310}) and (\ref{eq:312}), however, it follows that \textit{D} and \textit{T} transform as scalar fields in passing from one agn chart to another in the initial surface \textit{N} (i.e., that \(T = T'\) and \(D = D'\) in the regions of overlap). Thus \textit{D} and \textit{T} may be regarded as globally defined analytic functions on \textit{N}. From the Einstein equations \(R_{ab} = 0\), restricted to \textit{N} and reduced by means of \(\mathring{\varphi}_{,t} = k, \mathring{\mu}_{ab,3} = 0\) and \(\mathring{\beta}_{a,t3} = 0\), one can derive Eq.~(\ref{eq:303}) in any agn chart, which in turn implies the following differential equations for \textit{D} and \textit{T}:
\begin{equation}\label{eq:314}
D_{,3} = kD,\; T_{,3} = \frac{k}{2} T.
\end{equation}
The latter can be written more invariantly as \(\mathcal{L}_K D = kD\) and \(\mathcal{L}_K T = \frac{k}{2} T\) where \(\mathcal{L}_K\) represents Lie differentiation along the vector field \textit{K}.

Equations (\ref{eq:314}) show that (since \(k \neq 0\)) both \textit{D} and \textit{T} grow exponentially along the integral curves of \textit{K} in \textit{N}. However, the {Poincar\'{e}} recurrence argument of Sect.~\ref{subsec:application-poincare} has shown that each integral curve \(\gamma\) of \textit{K}, when followed arbitrarily far in either direction from any point \textit{p} on \(\gamma\), reapproaches \textit{p} arbitrarily closely. Since \textit{D} and \textit{T} are globally analytic (hence continuous) on \textit{N}, their values, when followed along \(\gamma\), would have to reapproach arbitrarily closely their values at \textit{p}. But this is clearly incompatible with their exponential growth along \(\gamma\). The only way to avoid this contradiction arises if \textit{D} and \textit{T} vanish globally on \textit{N}. We thus conclude that \(D = T = 0\) on \textit{N} and therefore, from the defining equations (\ref{eq:313}) and the fact that \(\mathring{\mu}_{ab}\) is positive definite, that
\begin{equation}\label{eq:315}
\mathring{h}_{ab} = \mathring{\mu}_{ab,t3} = 0
\end{equation}
on \textit{N}.

Now, computing the first \(t'\) derivatives of Eqs.~(\ref{eq:307}), restricting the results to the initial surface \(t = t' = 0\) and differentiating the resulting equations with respect to \(x^{3'}\) one finds, upon making use of Eqs.~(\ref{eq:301}), (\ref{eq:309}), and (\ref{eq:315}), that
\begin{equation}\label{eq:316}
\left.\frac{\partial^3 x^\alpha}{\partial x^{3'} \partial t' \partial t'}\right|_{t'=0} = 0
\end{equation}
whereas Eqs.~(\ref{eq:301}), (\ref{eq:302}) and (\ref{eq:315}) show that
\begin{equation}\label{eq:317}
\left.\vphantom{\frac{1}{2}}\left( g_{\alpha\beta, t3}\right)\right|_{t=0} = 0.
\end{equation}

We now proceed inductively to extend the above results to the case of time derivatives of arbitrarily high order. As an inductive hypothesis, suppose that, for some \(n \geq 1\) and for all \textit{k} such that \(0 \leq k \leq n\), we have
\begin{equation}\label{eq:318}
\begin{split}
\left.\left(\frac{\partial}{\partial x^3} \left(\frac{\partial^k g_{\alpha\beta}}{\partial t^k}\right)\right)\right|_{t=0} &= 0,\\
\left.\left(\frac{\partial}{\partial x^{3'}} \left(\frac{\partial^{k+1} x^\alpha}{\partial t'^{\,k+1}}\right)\right)\right|_{t'=0} &= 0,
\end{split}
\end{equation}
and recall that we also have
\begin{equation}\label{eq:319}
\left.\frac{\partial t}{\partial x^{3'}}\right|_{t'=0} = \left.\frac{\partial x^a}{\partial x^{3'}}\right|_{t'=0} = 0,\quad \left.\frac{\partial x^3}{\partial x^{3'}}\right|_{t'=0} = 1.
\end{equation}
Our aim is to prove that
\begin{equation}\label{eq:320}
\begin{split}
\left.\left(\frac{\partial}{\partial x^3} \left(\frac{\partial^{n+1} g_{\alpha\beta}}{\partial t^{n+1}}\right)\right)\right|_{t=0} &= 0,\\
\left.\left(\frac{\partial}{\partial x^{3'}} \left(\frac{\partial^{\, n+2} x^\alpha}{\partial t'^{n+2}}\right)\right)\right|_{t'=0} = 0.
\end{split}
\end{equation}

Note that the above imply that
\begin{equation}\label{eq:321}
\left.\left(\frac{\partial}{\partial x^3} \left(\frac{\partial^k g_{\alpha\beta}}{\partial x^{\gamma_1} \partial x^{\gamma_2} \ldots \partial x^{\gamma_k}}\right)\right)\right|_{t=0} = 0
\end{equation}
for all \(0 \leq k \leq n\) and for arbitrary \(\gamma_1, \gamma_2, \ldots , \gamma_k\). Furthermore, note that of the quantities \(\left.\left(\frac{\partial}{\partial x^3} \left(\frac{\partial^{n+1} g_{\alpha\beta}}{\partial x^{\gamma_1} \ldots \partial x^{\gamma_{n+1}}}\right)\right)\right|_{t=0}\), only \(\left.\left(\frac{\partial}{\partial x^3} \left(\frac{\partial^{n+1} g_{\alpha\beta}}{\partial t^{n+1}}\right)\right)\right|_{t=0}\), may be non-zero. Now differentiate the Einstein equation \(R_{t3} = 0\), \(n - 1\) times with respect to \textit{t} and set \(t = 0\) to derive an expression for \(\left.\left(\frac{\partial^{n+1}}{\partial t^{n+1}} \varphi\right)\right|_{t=0}\) in terms of \(x^3\)-invariant quantities. Differentiate the equation \(R_{tb} = 0\), \(n-1\) times with respect to \textit{t} and set \(t = 0\) to derive an expression for \(\left(\frac{\partial^{n+1}}{\partial t^{n+1}} \beta_b\right)|_{t=0}\), in terms of \(x^3\)-invariant quantities. Next, differentiate the equation \(R_{ab} = 0\), \textit{n} times with respect to \textit{t}, set \(t = 0\) and use the above results for \(\left.\left(\frac{\partial^{n+1}}{\partial t^{n+1}} \varphi\right)\right|_{t=0}\) and \(\left.\left(\frac{\partial^{n+1}}{\partial t^{n+1}} \beta_b\right)\right|_{t=0}\), together with those given in Eqs.~(\ref{eq:301}) and (\ref{eq:315}) to derive an equation of the form
\begin{equation}\label{eq:322}
\begin{split}
0 &= \left.\left(\frac{\partial^n}{\partial t^n} R_{ab}\right)\right|_{t=0}\\
 &= -\frac{\partial}{\partial x^3} \left.\left(\frac{\partial^{n+1}}{\partial t^{n+1}} \mu_{ab}\right)\right|_{t=0}\\
 &+ \binom{\hbox{positive}}{\hbox{constant}} \frac{\mathring{\varphi}_{,t}}{2} \left(\left.\frac{\partial^{n+1}}{\partial t^{n+1}} \mu_{ab}\right|_{t=0}\right)\\
 &+ \left\lbrace\hbox{terms independent of $x^3$}\right\rbrace.
\end{split}
\end{equation}
Differentiate this equation with respect to \(x^3\) to thus derive
\begin{equation}\label{eq:323}
\begin{split}
0 &= -\left(\left.\frac{\partial^{n+1}}{\partial t^{n+1}} \mu_{ab}\right|_{t=0}\right)_{,33}\\
 &+ \binom{\hbox{positive}}{\hbox{constant}} \frac{k}{2} \left(\left.\frac{\partial^{n+1}}{\partial t^{n+1}} \mu_{ab}\right|_{t=0}\right)_{,3}
\end{split}
\end{equation}
which holds in an arbitrary agn coordinate chart.

Now define
\begin{equation}\label{eq:324}
\begin{split}
D^{(n+1)} &\equiv \frac{\det{\left(\mathring{h}_{ab}^{(n+1)}\right)}}{\det{(\mathring{\mu}_{cd})}}\\
T^{(n+1)} &\equiv \mathring{\mu}^{ab} \mathring{h}_{ab}^{(n+1)}
\end{split}
\end{equation}
where \(\mathring{h}_{ab}^{(n+1)} \equiv \left.\left(\frac{\partial}{\partial x^3} \left(\frac{\partial^{n+1}}{\partial t^{n+1}} \mu_{ab}\right)\right)\right|_{t=0}\) so that Eq.~(\ref{eq:323}) becomes
\begin{equation}\label{eq:325}
0 = -\mathring{h}_{ab,3}^{(n+1)} + \binom{\hbox{positive}}{\hbox{constant}} \frac{k}{2} \mathring{h}_{ab}^{(n+1)}
\end{equation}
and \(D^{(n+1)}\) and \(T^{(n+1)}\) satisfy
\begin{equation}\label{eq:326}
\begin{split}
D^{(n+1)}_{,3} &= \binom{\hbox{positive}}{\hbox{constant}} kD^{(n+1)}\\
T^{(n+1)}_{,3} &= \binom{\hbox{positive}}{\hbox{constant}} \frac{k}{2} T^{(n+1)}
\end{split}
\end{equation}
in any agn coordinate chart. To extend the {Poincar\'{e}} recurrence argument to the quantities \(D^{(n+1)}\) and \(T^{(n+1)}\) we must first show that they are globally defined analytic functions on \textit{N}.

Differentiate the transformation equation
\begin{equation}\label{eq:327}
g_{a'b'} \equiv \mu_{a'b'} = \frac{\partial x^\alpha}{\partial x^{a'}} \frac{\partial x^\beta}{\partial x^{b'}} g_{\alpha\beta},
\end{equation}
\(n + 1\) times with respect to \(t'\), set \(t' = 0\) and differentiate the result with respect to \(x^{3'}\). Use the inductive hypothesis and the vanishing of \(\left.\left(\frac{\partial}{\partial x^3} \frac{\partial^{n+1}}{\partial t^{n+1}} \varphi\right)\right|_{t=0}\) and \(\left.\left(\frac{\partial}{\partial x^3} \frac{\partial^{n+1}}{\partial t^{n+1}} \beta_a\right)\right|_{t=0}\) to show that this calculation yields the covariance relation
\begin{equation}\label{eq:328}
\begin{split}
&\left.\left(\frac{\partial}{\partial x^{3'}} \frac{\partial^{n+1}}{\partial t'^{n+1}} \mu_{a'b'}\right)\right|_{t'=0}\\
&= \left.\left\lbrace\frac{\partial x^c}{\partial x^{a'}} \frac{\partial x^d}{\partial x^{b'}} \left(\frac{\partial}{\partial x^3} \frac{\partial^{n+1}}{\partial t^{n+1}} \mu_{cd}\right)\right\rbrace\right|_{t=0}
\end{split}
\end{equation}
From this and Eq.~(\ref{eq:310}) it follows that \(D^{(n+1)}\) and \(T^{(n+1)}\) transform as scalar fields in the overlap of agn charts in \textit{N} and thus that these quantities are globally defined analytic functions on \textit{N}. Equations (\ref{eq:326}) can thus be reexpressed in the invariant form
\begin{equation}\label{eq:329}
\begin{split}
\mathcal{L}_K D^{(n+1)} &= \binom{\hbox{positive}}{\hbox{constant}} kD^{(n+1)}\\
\mathcal{L}_K T^{(n+1)} &= \binom{\hbox{positive}}{\hbox{constant}} \frac{k}{2} T^{(n+1)}
\end{split}
\end{equation}
and show that \(D^{(n+1)}\) and \(T^{(n+1)}\) grow exponentially (unless they vanish) when followed along the integral curves of \textit{K} in \textit{N} (i.e., along the null generators of \textit{N}). Repeating the {Poincar\'{e}} recurrence argument given previously for \textit{D} and \textit{T} now yields a contradiction unless \(D^{(n+1)}\) and \(T^{(n+1)}\) vanish globally in \textit{N}. This in turn implies that
\begin{equation}\label{eq:330}
\left.\left(\frac{\partial}{\partial x^3} \frac{\partial^{n+1}}{\partial t^{n+1}} \mu_{ab}\right)\right|_{t=0} = 0
\end{equation}
in every agn chart on \textit{N} and, together with the results obtained above for the other metric components, shows that
\begin{equation}\label{eq:331}
\left.\left(\frac{\partial}{\partial x^3} \frac{\partial^{n+1}}{\partial t^{n+1}} g_{\alpha\beta}\right)\right|_{t=0} = 0
\end{equation}
in every such chart.

Applying the technique of the previous paragraph to the transformation equations for \(\varphi'\) and \(\beta'_a\) merely produces covariance relations for the quantities \(\left.\left(\frac{\partial}{\partial x^3} \left(\frac{\partial^{n+1}}{\partial t^{n+1}} \varphi\right)\right)\right|_{t=0}\) and \(\left.\left(\frac{\partial}{\partial x^3} \left(\frac{\partial^{n+1}}{\partial t^{n+1}} \beta_a\right)\right)\right|_{t=0}\) which are consistent with the (already established) vanishing of these quantities in every agn chart. To complete the inductive proof, we differentiate the remaining transformation equations (\ref{eq:307}) \(n + 1\) times with respect to \(t'\), set \(t' = 0\), use the inductive hypothesis and the new results summarized in Eq.~(\ref{eq:331}) to show that
\begin{equation}\label{eq:332}
\left.\left(\frac{\partial}{\partial x^{3'}} \frac{\partial^{n+2}}{\partial t'^{\, n+2}} x^\alpha\right)\right|_{t'=0} = 0.
\end{equation}
This result, together with that of Eq.~(\ref{eq:331}), completes the proof by induction.

It follows from the analyticity of \textit{g} and the inductive proof given above that \(\left(\frac{\partial}{\partial x^3} g_{\alpha\beta}\right)\) vanishes throughout any agn coordinate chart and thus that \(Y \equiv \frac{\partial}{\partial x^3}\) is a (locally defined) analytic Killing field throughout the given chart. In the region of overlap of any two such charts we have the two locally defined Killing fields \(Y = \frac{\partial}{\partial x^3}\) and \(Y' = \frac{\partial}{\partial x^{3'}}\) and we wish to show that, in fact, they coincide. By construction both \textit{Y} and \(Y'\) coincide with \textit{K} on their appropriate domains of definition within the null surface \textit{N}. Therefore \(X \equiv Y' - Y\) is an analytic Killing field of \textit{g} defined locally on the region of overlap of the two charts which vanishes on the intersection of this region with the null surface \textit{N}. This implies that \textit{X} vanishes throughout its domain of definition, however, since the Killing equations
\begin{equation}\label{eq:333}
X_{\mu,t} + X_{t,\mu} - 2 {}^{(4)}\!\Gamma_{\mu t}^\nu X_\nu = 0
\end{equation}
determine \textit{X} uniquely from data \(X|_{t=0}\) (in the analytic case) and have only the trivial solution \(X = 0\) if \(X|_{t=0} = 0\).

It follows from the above that there exists a unique analytic Killing field \textit{Y}, globally defined on a full neighborhood of \textit{N} in \(({}^{(4)}\!V, g)\) which, when restricted to \textit{N}, coincides with the vector field \textit{K} and this is tangent to the null generators of \textit{N}. In fact, one can prove that \textit{Y} extends to a Killing field defined throughout the maximal Cauchy development of the globally hyperbolic region of \(({}^{(4)}\!V, g)\) whose Cauchy horizon is \textit{N}. The techniques for proving this were discussed at the end of section III of Ref.~\cite{Moncrief:1983} and need not be repeated here. One can also show, by a straightforward computation that
\begin{equation}\label{eq:334}
\left.\left\lbrace Y^\beta {}^{(4)}\!\nabla_\beta Y^\alpha + \frac{k}{2} Y^\alpha\right\rbrace\right|_N = 0
\end{equation}
which suggests that the constant \(\left(-\frac{k}{2}\right)\) is the analogue, for cosmological Cauchy horizons, of the \textit{surface gravity} defined for stationary black hole event horizons \cite{Moncrief:2008,Hawking:1973}.

We have thus proven:

\begin{theorem}
  Let \(({}^{(4)}\!V,g)\) be a real analytic, time orientable, vacuum spacetime which admits a compact, connected Cauchy horizon \textit{N} that separates \(({}^{(4)}\!V,g)\) into open Lorentzian submanifolds \(({}^{(4)}\!V_+,g_+)\) and \(({}^{(4)}\!V_-,g_-)\) of which one is globally hyperbolic and the other acausal. Assume that \textit{N} is realized as a level set of some analytic function \(\tau: {}^{(4)}\!V \rightarrow \mathbb{R}\) having no critical points in a neighborhood of \textit{N}. The vector field \({}^{(4)}\!X := \mathrm{grad}_g\tau\) will therefore be non-vanishing on this neighborhood, null on the hypersurface \textit{N} and thus tangent to its null geodesic generators and will naturally induce (by restriction of \({}^{(4)}\!X\) to \textit{N}) a corresponding tangent vector field \textit{X} on the Cauchy horizon itself.

  In the cases referred to here as `non-ergodic' the null generators of \textit{N} are either closed curves or densely fill 2-tori embedded in \textit{N} and every such generator is either complete in both the directions of \textit{X} and \(-X\) (the `degenerate' case) or else every generator is incomplete in one direction (say that of \textit{X}) and complete in the opposite direction (the `non-degenerate' case).

  Compact, non-degenerate, non-ergodic Cauchy horizons in analytic, vacuum spacetimes \(({}^{(4)}\!V,g)\) are Killing horizons in that there always exists a non-trivial, analytic Killing field \textit{Y}, globally defined on a full neighborhood of the horizon manifold \(N \subset ({}^{(4)}\!V,g)\) which, when restricted to \textit{N}, is everywhere tangent to the null generators of this hypersurface. \textit{Y} extends (at least smoothly) to a Killing field defined throughout the maximal Cauchy development of the globally hyperbolic region of \(({}^{(4)}\!V,g)\) whose Cauchy horizon is \textit{N}.
\end{theorem}

By applying the results of our earlier work (cf. Ref.~\cite{Isenberg:1992} and Sect. VIII of Ref.~\cite{Moncrief:2008}) it is straightforward to prove that if the null generators of \textit{N}, to which the horizon generating Killing field \textit{Y} is tangent, are not all closed curves then the globally hyperbolic region of \(({}^{(4)}\!V,g)\) necessarily admits at least one additional, non-trivial Killing field. This additional Killing field commutes with \textit{Y} so that the full isometry group of this (globally hyperbolic) spacetime includes a 2-dimensional toral action.

Thus whereas non-degenerate Cauchy horizons having only closed (null geodesic) generators are, in a geometrical sense, less `general' than those admitting non-closed generators they are, nevertheless, far less constrained analytically in that they can bound (analytic, vacuum) globally hyperbolic spacetimes having only one-dimensional  isometry groups. Furthermore, if our conjecture for the (non-degenerate) ergodic case is correct then the solution set for these is much smaller still, consisting uniquely of certain `irrational' compactifications of the flat Kasner spacetime.

Finally, though we could only rule out the existence of \textit{degenerate} (compact, analytic) Cauchy horizons in some (closed-orbit) special cases \cite{Moncrief:1983} we conjecture that such horizons do not exist at all. 

\section*{Acknowlegements}
The authors would especially like to thank the Universit\'{e} Paris VI, the Institut des Hautes \'{E}tudes Scientifique in Bures-sur-Yvette, France, the Albert Einstein Institute in Golm, Germany, the Erwin Schr\"{o}dinger Institute in Vienna, Austria, the Isaac Newton Institute in Cambridge, England, the Kavli Institute for Theoretical Physics in Santa Barbara, California, the Mathematical Sciences Research Institute in Berkeley, California and the Mittag-Leffler Institute in Djursholm, Sweden for their warm hospitality and generous support during visits over the years when portions of this research were carried out. We are particularly grateful to Maxim Kontsevich for providing the theorem quoted in the Appendix. We also thank Lars Andersson for valuable discussions.

\appendix
\section*{Appendix}
To show that each of our embedded 2-tori admits an analytic foliation, with closed leaves, that is everywhere transverse to the (nowhere vanishing) flow field \textit{X} it would suffice to prove that it always admits a closed, analytic one-form \(\lambda\) with integral periods such that \(\lambda (X) = \lambda_a X^a > 0\) everywhere on the given torus. The closure of \(\lambda\) ensures that, locally, it is expressible as \(\lambda = d\mu\) for some analytic function \(\mu\) the level curves of which locally define the leaves of the desired foliation. That these leaves all close, globally, is ensured by the integrality of the periods of \(\lambda\) whereas their transversality to \textit{X} corresponds simply to the condition that \(\lambda (X) > 0\).

The following proof that such a \(\lambda\) always exists is due to M.~Kontsevich who kindly provided it to us in response to a question about a somewhat related theorem of Kolmogorov's. Note that analyticity is not needed for some of the intermediate steps of Kontsevich's argument but that it will be `reinstated' during the final stage of the construction.

First choose a smooth Siegel curve \(\tilde{\Gamma}\) that is closed, non-self-intersecting and everywhere transverse to the flow of \textit{X}. The existence of such curves follows from a standard argument which is given, for example, in \cite{Cornfeld:1982} together with a discussion of some of their fundamental properties. The aim will be to construct an analytic foliation whose leaves are each homotopic to  \(\tilde{\Gamma}\) (and transversal to \textit{X}). By translating \(\tilde{\Gamma}\) along the flow generated by \textit{X} one can produce a curve, homotopic to \(\tilde{\Gamma}\), that passes through any particular point of the  given torus and that is, of course, also transversal to \textit{X}.

Any one of such Siegel curves, \(\Gamma\), can by systematically `thickened' to yield a smooth `ribbon', \(r_\Gamma\), diffeomorphic to \(\Gamma \times I_\Gamma \approx \mathbf{S}^1 \times I_\Gamma\) where \(I_\Gamma\) is an open interval. Coordinatize this ribbon by choosing an `angle' coordinate \(\theta_\Gamma\) along \(\Gamma\), with \(\theta_\Gamma \in [0, 2\pi)\), and letting \textit{t} be the flow parameter along the transversal flow generated by \textit{X}, with \(t \in I_\Gamma := (-\epsilon_\Gamma, \epsilon_\Gamma)\) for some \(\epsilon_\Gamma > 0\), taking \(t = 0\) to correspond to the given `source curve' \(\Gamma\).

Now define a \textit{smooth} one-form \(\alpha_\Gamma\) on the torus by setting \(\alpha_\Gamma = 0\) on the complement of the ribbon \(r_\Gamma\) but taking \(\alpha_\Gamma = d\mu_\Gamma\) within the ribbon where \(\mu_\Gamma\) is a smooth function of \textit{t} alone (i.e., independent of \(\theta_\Gamma\)) that smoothly and monotonically interpolates between the value 0 for \(t \in (-\epsilon_\Gamma, -\epsilon_\Gamma/2)\) and the value 1 for \(t \in (\epsilon_\Gamma/2, \epsilon_\Gamma)\) with derivative satisfying \(\frac{\partial}{\partial t} \mu_\Gamma \geq 0\) for \(t \in (-\epsilon_\Gamma, \epsilon_\Gamma)\) and \(\frac{\partial}{\partial t} \mu_{\Gamma} > 0\) for \(t \in (-\epsilon_\Gamma/2, \epsilon_\Gamma/2)\). The one-form \(\alpha_\Gamma\) so-constructed will be closed, have integral periods and satisfy \(\alpha_\Gamma (X) \geq 0\) everywhere on the chosen torus.

In  view of the compactness of the torus a finite collection, \(\lbrace r_{\Gamma_i}; i = 1, \ldots, k\rbrace\), of such ribbons, together with their associated closed one-forms, \(\lbrace\alpha_{\Gamma_i}; i = 1, \ldots, k\rbrace\) will suffice to cover the torus in such a way that
\begin{equation*}
\alpha := \sum_{i=1}^k \alpha_{\Gamma_i}
\end{equation*}
satisfies \(d\alpha = 0, \alpha (X) > 0\) everywhere and has integral periods (since each of the \(\alpha_{\Gamma_i}\) does). It will not however be analytic since none of the individual \(\alpha_{\Gamma_i}\)'s are more than smooth.

Taking, however, a Hodge decomposition of \(\alpha\) with respect to an analytic (Riemannian) metric on the torus will result in
\begin{equation*}
\alpha = h + d\sigma
\end{equation*}
where \textit{h} is harmonic and thus analytic but where the function \(\sigma\) is only smooth. The integral periods of \(\alpha\) will all be `carried' by \textit{h} since of course those of \(d\sigma\) all vanish. Now, however, since the condition \(\alpha (X) > 0\) is \textit{open} one can always preserve it by approximating \(\sigma\) with an analytic function \(\omega\). Thus defining
\begin{equation*}
  \lambda = h + d\omega
\end{equation*}
one arrives at a closed, analytic one-form with integral periods that globally satisfies the transversality condition \(\lambda (X) > 0\) and thereby determines an analytic foliation of the torus of the type desired. 

\bibliographystyle{unsrt} 
\bibliography{paper_refs}   

\end{document}